\documentclass[preprint,prd,showpacs]{revtex4}
\usepackage{amsmath,epsfig}
\usepackage{amsfonts}
\usepackage{amssymb}
\usepackage{graphicx}
\usepackage[cmtip,arrow]{xy}
\usepackage{url}

\usepackage{amssymb,amsopn}

\newtheorem{tw}{Theorem}

\newtheorem{co}{Corollary}

\newtheorem{ex}{Example}

\renewcommand{\theequation} {\arabic{section}.\arabic{equation}}

\newcounter{orange}
\renewcommand{\theorange}{\alph{orange}}

\begin{document}
\title{Uncertainty relations in quantum optics. Is the photon intelligent?\footnote{In memory of Jerzy F. Pleba\'{n}ski}}

\author{Maciej Przanowski}\email{maciej.przanowski@p.lodz.pl}
\affiliation{Institute of Physics, {\L}\'{o}d\'{z} University of
Technology, Wolczanska 219, 90-924 {\L}\'od\'{z}, Poland.}

\author{Hugo Garc\'{\i}a-Compe\'an}
\email{compean@fis.cinvestav.mx} \affiliation{Departamento
de F\'{\i}sica, Centro de Investigaci\'on y de Estudios
Avanzados del IPN\\ P.O. Box 14-740, 07000 M\'exico D.F., M\'exico.}

\author{Jaromir Tosiek}\email{jaromir.tosiek@p.lodz.pl}
\affiliation{Institute of Physics, {\L}\'{o}d\'{z} University of
Technology, Wolczanska 219, 90-924 {\L}\'od\'{z}, Poland.}

\author{Francisco J. Turrubiates}\email{fturrub@esfm.ipn.mx}
\affiliation{Departamento de F\'{\i}sica, Escuela Superior de F\'{\i}sica
y Matem\'aticas del IPN\\ Unidad Adolfo
L\'opez Mateos, Edificio 9, 07738  M\'exico D.F., M\'exico.}
\date{\today}

\begin{abstract}
The Robertson -- Schr\"{o}dinger, Heisenberg -- Robertson and Trifonov
uncertainty relations for arbitrary two functions $f_{1}$ and $f_{2}$
depending on the quantum phase and the number of photons
respectively, are given. Intelligent states and  states which
minimize locally the product of uncertainties $(\Delta
f_{1})^{2}\cdot (\Delta f_{2})^{2}$ or the sum $(\Delta
f_{1})^{2}+(\Delta f_{2})^{2}$ are investigated for the cases
$f_{1}=\phi,\exp{(i\phi)}, \exp{(-i\phi)}, \cos{\phi}, \sin{\phi}$
and $f_{2}=n$.
\end{abstract}

\vskip -1truecm

\vskip 1truecm

\maketitle

\vskip -1.3truecm
\newpage

\setcounter{equation}{0}

\section{Introduction}

The principle of uncertainty for the position and the momentum formulated by
Werner Heisenberg in 1927 \cite{heisenberg} is one of the most
fundamental results not only in quantum mechanics but also in all the
physics. It is not strange that this principle has attracted a
great deal of interest. First, it was proved with the mathematical
rigour by E.H. Kennard \cite{kennard} and H. Weyl \cite{weyl}. Then
the original uncertainty relations for canonical variables were
generalized to the case of any two observables by H.P. Robertson
\cite{robertson1,robertson2} and E. Schr\"{o}dinger
\cite{schrodinger} and also to the case of any number of observables
by H.P. Robertson \cite{robertson3}. The Robertson approach was
thoroughly analysed and developed by D.A. Trifonov in the series of
papers
\cite{trifonov1,trifonov2,trifonov3,trifonov4,trifonov5,trifonov6,trifonov7,trifonov8}
especially in the context of intelligent, coherent and squeezed
states in order to extend the notion of these states on the cases
when the generalized uncertainty relations are considered (see also
\cite{dodonov}). 

As we remember the (\textit{usual})
\textit{coherent state} is a state which minimizes the Heisenberg
uncertainty relation for the canonical variables
$\widehat{q}=\frac{1}{\sqrt{2}}(\widehat{a}+\widehat{a}^{\dag})$ and
$\widehat{p}=\frac{i}{\sqrt{2}}(\widehat{a}^{\dag}-\widehat{a})$
with $\Delta q = \Delta p$, where $\Delta q$ and $\Delta p$ stand
for the uncertainties of $q$ and $p$, respectively
\cite{klauder,zang,tanas,kryszewski}.
 The  (\textit{usual})
\textit{squeezed states} minimize the Heisenberg uncertainty
relation for $\widehat{q}$ and $\widehat{p}$ but $\Delta q \neq
\Delta p$ \cite{tanas,kryszewski,IP1,pleban,yuen,loudon}. 
The coherent
and squeezed states are called the \textit{intelligent states} and they
can be defined as the states which give the strict equality in the
Heisenberg uncertainty relation for $\widehat{q}$ and $\widehat{p}$
or, equivalently, as the states which minimize globally  the product $\Delta q
\cdot \Delta p$. It is obvious that in the case of a generalized
uncertainty relation for arbitrary observables when the right hand
side of the respective inequality depends on the state, these two definitions of an intelligent state given above are in general not
equivalent. According to the convention accepted in the previous works
\cite{trifonov2,trifonov3,trifonov4,aragone,vaccaro,PT,smithhey,brif1,pegg1}
we refer to a generalized intelligent state as the one which satisfies
the equality in the respective generalized uncertainty relation.

It is worthwhile mentioning that the Robertson approach to the uncertainty relations
and also the concept of the generalized intelligent
states can be carried over to the deformation quantization formalism \cite{PT,curthright}.

In the present paper we investigate  various generalized
uncertainty relations in quantum optics for quantities depending on a
phase and a number of photons. Then we study the corresponding
intelligent states and the states which minimize (locally) the
product or sum of uncertainties. The interest in this issue starts
with the works of P. Carruthers and M.M. Nieto
\cite{carruthers1,carruthers2} and R. Jackiw \cite{jackiw}. Then the
problem has been explored by many authors
\cite{vaccaro,smithhey,brif1,garrison,levy,yamamoto,luks1,luks2,brif2,mendas,luo,skagerstam,scharat, shapiro}.
Since the generalized uncertainty relations for  functions of the
phase and the number of photons depend on the quantum state, the
usual procedure in searching for the states minimizing these
relations consists in modifying the respective inequality to obtain
a constant parameter on the right hand side of this inequality.
Then one minimizes such an inequality
\cite{carruthers1,carruthers2,jackiw,luo}. 

In our paper we are going
to use another technique. We simply leave the generalized
uncertainty relations as they stand and we ask for the states which
give the \textit{local minimum} of these relations. To the best of our
knowledge such a question has not been yet considered in quantum
optics but it has been widely investigated for the `angular momentum
- angle position' uncertainty relation in \cite{pegg1,pereira}. In
fact our work is strongly motivated by Ref. \cite{pegg1}.

  As we will
see, the results obtained in our paper are drastically different from
the corresponding results in \cite{pegg1}, since in contrast to the
case of angular momentum operator $\widehat{L}_{z},$ which has the
eigenvalues $n\hbar$, $n=0,\pm 1, \pm 2, \ldots ,$ the photon number
operator $\widehat{n}$ admits only the eigenvalues $n=0,1,2, \ldots $
Moreover, if one is going to study the `photon number -- phase
function' uncertainty relations, then he/she must first decide on the
formalism in which he/she considers the quantum phase. 
He/she may deal within the Susskind -- Glogower formalism
\cite{susskind,tanas,carruthers2,jackiw}, the Garrison -- Wong formalism
\cite{garrison} or  he/she can apply the Pegg -- Barnett approach
\cite{pegg2,pegg3,barnett}, which is equivalent to the POV -- measure
approach \cite{busch,shapiro} and to the formalism based on
extending the Fock space to the Hilbert space $L^{2}(S^{1})$
\cite{przan1,przan2} (see also \cite{perinova} and references therein). In the present paper we employ the results of
\cite{przan1,przan2} and  our analysis is consistent with the
celebrated Pegg -- Barnett approach to the quantum phase.

The paper is organized as follows. In Sec. 2 we first recall some
results of \cite{przan1,przan2} and then we quote the Robertson,
Hadamard -- Robertson and Trifonov theorems. In the next step, using
these theorems we derive the Robertson -- Schr\"{o}dinger,
Heisenberg -- Robertson and Trifonov uncertainty relations for any two
functions $f_{1}=f_{1}(\phi)$ and $f_{2}=f_{2}(n)$ depending on the
phase $\phi$ and the number of photons $n$. Finally, we employ these
general results  to the case of the phase and the number of
photons (Example \ref{ex2.1}). 

The number -- phase Robertson -- Schr\"{o}dinger,
Heisenberg -- Robertson and Trifonov intelligent states are found
(Theorem \ref{t2.1}). Intelligent states for arbitrary $f_{1}$ and $f_{2}$
are investigated in Sec 3. In particular the cases:
$f_{1}=\exp({i\phi})$ and $f_{2}=n$ (Example \ref{ex3.1}),
$f_{1}=\exp{(-i\phi)}$ and $f_{2}=n$ (Example \ref{ex3.2}), and finally
$f_{1}=\cos{\phi}$ ($f_{1}=\sin{\phi}$) and $f_{2}=n$ (Example \ref{ex3.3})
are considered in all details.

  Section 4 is devoted to searching for
the states which \textit{minimize locally} the product of
uncertainties $(\Delta f_{1})^{2}\cdot (\Delta f_{2})^{2}$. The
general equation (see Eq. (\ref{generaleqn45})) defining these
states has been found and the cases of $f_{1}=\exp{(-i\phi)}$
(Example \ref{ex4.1}) and $f_{1}=\phi$ (Example \ref{ex4.2}) with $f_{2}=n$ (in both
cases), have been studied in detail. The same is done in the next
section,  where the states \textit{minimizing locally} the sum of
uncertainties $(\Delta f_{1})^{2}+(\Delta f_{2})^{2}$ are
investigated. Concluding remarks  end our paper.

  It is
worth reminding that our results are valid within the Pegg -- Barnett
approach to quantum phase as well as within the POV -- measure approach
or within the formalism based on extending the Fock space to the
Hilbert space $L^{2}(S^{1})$ \cite{przan1,przan2}. However, the
Susskind -- Glogower or Garrison -- Wong formalisms lead, in general, to
different results than the ones presented here.

We devote this modest work to the memory of our Teacher and Friend
Jerzy F. Pleba\'{n}ski into the tenth anniversary of his death. Professor Pleba\'{n}ski was not only a great relativist but he was also the first, with Leopold Infeld, who considered already in the years 1954 -- 1955 the squeezed states and the squeeze operator for a harmonic oscillator \cite{IP1, pleban}. Hardly anyone knows this fact. We have found  comments  about the mentioned publications  in \cite{miranowicz, octavio}.


\section{Uncertainty Relations}

In the recent work \cite{przan2} it has been argued that given a
number -- phase function $f=f(\phi,n), \, -\pi \leq \phi < \pi, \, n=0,1, \ldots,$ the average value of this
function in a state defined by a density operator $\widehat{\rho}$
is given by
\begin{equation}\label{21}
  \langle f(\phi,n)\rangle = \sum_{n=0}^{\infty}\int_{-\pi}^{\pi}f(\phi,n)\rho_{W}(\phi,n)d\phi = {\rm Tr} \{\widehat{f}(\phi,n)\widehat{\rho}\}= \langle \widehat{f}(\phi,n)\rangle,
\end{equation}
where $\rho_W(\phi,n)$ is the number -- phase Wigner function for the
state $\widehat{\rho}$
\begin{equation}\label{22}
  \rho_{W}(\phi,n)= {\rm Re} \left \{\langle \phi | \widehat{\rho}| n\rangle \langle n|\phi \rangle \right \}.
\end{equation}
The operator $\widehat{f}(\phi,n)$ reads
\begin{equation}\label{23}
  \widehat{f}(\phi,n)=\frac{1}{2}\sum_{n=0}^{\infty}\int_{-\pi}^{\pi}f(\phi,n) \cdot
  \Big( |n\rangle\langle n |\phi\rangle\langle\phi| +
  |\phi\rangle\langle \phi |n\rangle\langle\ n| \Big) d\phi.
\end{equation}
The vectors $|\phi\rangle$ and $|n\rangle$ stand for the phase state vector \cite{london,carruthers2,brif1,brif2,luis,biatynicka,perinova,tanas,przan2}
 \begin{equation}\label{24}
  |\phi\rangle=\frac{1}{\sqrt{2\pi}}\sum_{n=0}^{\infty}\exp{(i n \phi)} |n\rangle
\end{equation}
and the normalized eigenvector of the number operator $\widehat{n}$
\begin{eqnarray}\label{25}
  \widehat{n}|n\rangle &=& n |n\rangle ,  \ \ \ \ \ n=0,1,\ldots  \nonumber \\
  \langle n'|n\rangle &=& \delta_{n' n}
\end{eqnarray}
respectively.

\noindent
\textbf{Remark} In the present work we do not use the
`underbar notation' $\underline{\rho_{W}}, \underline{|n\rangle},
\underline{|\phi\rangle}, \ldots$ etc. which has been employed in
\cite{przan1,przan2}; we also omit the kernel symbol ${\cal
K}_{S}.$

In particular if the function $f$ is independent of the photon number $n$ i.e. $f=f(\phi)$ then the formula (\ref{21}) reduces to
\begin{equation}\label{26}
  \langle f(\phi)\rangle=\int_{-\pi}^{\pi} f(\phi) \langle \phi|\widehat{\rho}|\phi \rangle d\phi
  = {\rm Tr} \{\widehat{f}(\phi) \widehat{\rho} \}
  = \langle \widehat{f}(\phi)\rangle
\end{equation}
where $\widehat{f}(\phi)$, according to (\ref{23}), is given by
\begin{equation}\label{27}
  \widehat{f}(\phi)=\int_{-\pi}^{\pi} f(\phi)|\phi\rangle \langle \phi| d\phi.
\end{equation}
Recall that the set of vectors $\{|\phi\rangle\}_{-\pi}^{\pi}$ is not orthogonal
\begin{equation}\label{28}
 \langle \phi |\phi'\rangle \nsim \delta(\phi-\phi')
\end{equation}
but still it gives a resolution of the identity operator
\begin{equation}\label{29}
  \int_{-\pi}^{\pi} |\phi\rangle\langle \phi| d\phi = \widehat{1}.
\end{equation}
If the function $f$ is independent of the phase $\phi$ i.e.  $f=f(n)$ then (\ref{21}) takes the form
\begin{equation}\label{210}
  \langle f(n)\rangle = \sum_{n=0}^{\infty} f(n) \langle n | \widehat{\rho}|n \rangle = {\rm Tr} \{\widehat{f}(n)\widehat{\rho}\} = {\rm Tr} \{ f(\widehat{n})\widehat{\rho} \}=\langle f(\widehat{n})\rangle.
\end{equation}
Note that because of (\ref{28}) if $f=f(\phi)$ and $g=g(\phi)$ then
\begin{equation}\label{211}
  \widehat{fg}(\phi)\neq \widehat{f}(\phi)\cdot\widehat{g}(\phi)
\end{equation}
in general, and consequently it may happen that
\begin{equation}\label{212}
  \langle f(\phi) g(\phi)\rangle=\langle \widehat{fg}(\phi)\rangle \neq \langle \widehat{f}(\phi) \cdot \widehat{g}(\phi)\rangle.
\end{equation}

In particular
\begin{equation}\label{213}
  \langle \phi^{k}\rangle = \langle \widehat{\phi^{k}} \rangle \neq \langle (\widehat{\phi})^{k} \rangle,
\end{equation}
where
\begin{equation}\label{214}
  \widehat{\phi}=-i\sum^{\infty}_{j,l=0}\frac{(-1)^{j-l}}{j-l}|j\rangle \langle l|     \,\,\,\,\, \ \
    \textrm{with} \,\, \ \ \ \ \ j \neq l
\end{equation}
is the self -- adjoint \textit{Garrison -- Wong phase operator} \cite{garrison,tanas,busch,przan1}.

On the contrary, if $f=f(n)$ and $g=g(n)$ then
\begin{equation}\label{215}
  \widehat{fg}(n)=\widehat{f}(n)\cdot \widehat{g}(n)=f(\widehat{n})\cdot g(\widehat{n}).
\end{equation}
It can be easily proved that the average value (\ref{26}) is equal
exactly to the one calculated within the celebrated Pegg -- Barnett
formalism
\cite{pegg2,pegg3,barnett,busch,shapiro,przan1,przan2,luis}. Then
the formulae (\ref{211}), (\ref{212}) and (\ref{213}) show that the
Pegg -- Barnett formalism predicts different experimental results than
the formalism which assumes that the quantum phase is defined by the
Garrison -- Wong operator (\ref{214}) and also different than the
results predicted by the famous Susskind -- Glogower formalism, where
the quantum phase is given by two hermitian (self -- adjoint) operators
$\widehat{\cos \phi}$ and $\widehat{\sin \phi}$
\cite{gerhardt,nieto,lyrich,gerry,gantsog}.

Consequently, one expects that all those three approaches lead also
to different uncertainty relations. In this work we study the
uncertainty relations which follow from the Pegg -- Barnett approach.

Let
\begin{equation}\label{216}
  \widehat{\rho}=|\psi \rangle \langle \psi |, \,\,\,\,\, \langle \psi |\psi \rangle=1
\end{equation}
be the density operator for a pure state $|\psi\rangle$. Given any two
complex functions $f_{1}=f_{1}(\phi)$ and $f_{2}=f_{2}(n)$ one
quickly gets from $(\ref{26})$ and $(\ref{210})$ their average values
\setcounter{orange}{1}
\renewcommand{\theequation} {\arabic{section}.\arabic{equation}\theorange}
\begin{equation}\label{217a}
  \langle f_{1}(\phi) \rangle = \int_{-\pi}^{\pi} \psi^{*}(\phi)f_{1}(\phi)\psi(\phi) d\phi,
\end{equation}
\addtocounter{orange}{1}
\addtocounter{equation}{-1}
\begin{equation}\label{217b}
  \langle f_{2}(n) \rangle = \int_{-\pi}^{\pi} \psi^{*}(\phi)f_{2}\left(i\frac{\partial}{\partial\phi}\right)\psi(\phi)
  d\phi,
\end{equation}
\renewcommand{\theequation} {\arabic{section}.\arabic{equation}}
where, with $|\phi\rangle$ defined by (\ref{24}), the function
\begin{equation}\label{218}
  \psi(\phi)=\langle \phi |\psi \rangle
\end{equation}
is the \textit{wave function in the phase representation}.

Introduce two operators acting in the space of such wave functions 
\setcounter{orange}{1}
\renewcommand{\theequation} {\arabic{section}.\arabic{equation}\theorange}
\begin{equation}\label{219a}
  \widehat{F}_{1}\psi(\phi) := f_{1}(\phi)\psi(\phi),
\end{equation}
\addtocounter{orange}{1}
\addtocounter{equation}{-1}
\begin{equation}\label{219b}
  \widehat{F}_{2}\psi(\phi) := f_{2}\left(i\frac{\partial}{\partial\phi}\right)\psi(\phi)
\end{equation}
\renewcommand{\theequation} {\arabic{section}.\arabic{equation}}
and then the  following two operators
\setcounter{orange}{1}
\renewcommand{\theequation} {\arabic{section}.\arabic{equation}\theorange}
\begin{equation}\label{220a}
  \delta\widehat{F}_{1} := \widehat{F}_{1}-\langle \widehat{F}_{1} \rangle = \widehat{F}_{1}-\langle
  f_{1}(\phi)\rangle,
\end{equation}
\addtocounter{orange}{1}
\addtocounter{equation}{-1}
\begin{equation}\label{220b}
  \delta\widehat{F}_{2} := \widehat{F}_{2}-\langle \widehat{F}_{2} \rangle = \widehat{F}_{2}-\langle
  f_{2}(n)\rangle .
\end{equation}
\renewcommand{\theequation} {\arabic{section}.\arabic{equation}}

Let us define now the $2\times2$ Hermitian matrix
\begin{equation}\label{221}
  F_{\mu\nu}:=\int_{-\pi}^{\pi} \big(\delta\widehat{F}_{\mu} \psi(\phi)\big)^{*}\delta\widehat{F}_{\nu}\psi(\phi)d\phi \,\,\,\,\, \ \  \textrm{with} \,\, \ \
  \mu,\nu=1,2.
\end{equation}
It is obvious that the Hermitian form ${\cal F}$
\begin{equation}\label{222}
  {\cal F}:\mathbb{C}^{2}\times\mathbb{C}^{2}\ni(x,y)\longmapsto {\cal F}(x,y)=\sum_{\mu,\nu = 1}^{2}F_{\mu\nu}x_{\mu}^{*}y_{\nu}\in\mathbb{C}
\end{equation}
is \textit{positive semi -- definite} i.e. ${\cal F}(x,x)\geq 0 \,\, \forall \,\, x \in \mathbb{C}^{2}$. We split the matrix ($F_{\mu\nu}$) into two matrices
\begin{eqnarray}\label{223}
  (F_{\mu\nu}) &=& (a_{\mu\nu})+i(b_{\mu\nu}), \nonumber \\
    a_{\mu\nu} &:=& \frac{1}{2}(F_{\mu\nu} + F_{\nu\mu})=a_{\nu\mu}=a_{\mu\nu}^{*}, \nonumber \\
    b_{\mu\nu} &:=& \frac{1}{2i}(F_{\mu\nu} -
    F_{\nu\mu})=-b_{\nu\mu}=b_{\mu\nu}^{*}.
\end{eqnarray}
The matrix $(a_{\mu\nu})$ is real symmetric and $(b_{\mu\nu})$ is real anti -- symmetric therefore Hermitian
and anti -- Hermitian respectively.

Uncertainty relations for $f_{1}(\phi)$ and $f_{2}(n)$ follow
directly from the general theorems known as: the \textit{Robertson
theorem} \cite{robertson3,trifonov7,trifonov8,PT}, the
\textit{Hadamard -- Robertson theorem}
\cite{robertson3,trifonov2,trifonov3,trifonov8,PT} and the
\textit{Trifonov theorem} \cite{trifonov8,PT}.

In our present case the Robertson theorem states that
\begin{equation}\label{224}
  \det(a_{\mu\nu}) \geq \det(b_{\mu\nu}).
\end{equation}
Then the Hadamard -- Robertson theorem shows that
\begin{equation}\label{225}
  a_{11}a_{22} \geq \det(b_{\mu\nu})
\end{equation}
and the Trifonov theorem reduces now to
\begin{equation}\label{226}
  a_{11} + a_{22} \geq 2 |b_{12}|.
\end{equation}
Notice that (\ref{224}) leads to a stronger estimation of $\det(b_{\mu\nu})$ then (\ref{225}).

Employing (\ref{221}) and (\ref{223}) with (\ref{219a}), (\ref{219b}), (\ref{220a}) and (\ref{220b}) after some elementary algebraic manipulations one gets
from (\ref{224}) the following \textit{Robertson -- Schr\"{o}dinger uncertainty relation}
\begin{eqnarray}
\label{227}
 (\Delta f_{1})^{2}\cdot (\Delta f_{2})^{2} &\geq& \left|\int_{-\pi}^{\pi}(\delta\widehat{F}_{1}\psi(\phi))^{*} \delta\widehat{F}_{2}\psi(\phi) d\phi \right|^{2}
 \nonumber \\
   &=& \left|\int_{-\pi}^{\pi} \psi^{*}(\phi)f_{1}^{*}(\phi)f_{2}\left(i\frac{\partial}{\partial\phi}\right)\psi(\phi) d\phi - \langle f_{1}(\phi)\rangle^{*}\langle f_{2}(n)\rangle
   \right|^{2},
\end{eqnarray}
where
\begin{equation}
\label{228}
  (\Delta f_{\mu})^{2} := \int_{-\pi}^{\pi}\left(\delta\widehat{F}_{\mu}\psi(\phi)\right)^{*} \delta\widehat{F}_{\mu}\psi(\phi) d\phi 
   = F_{\mu\mu}=a_{\mu\mu}, \,\,\,\,\,\, \mu=1,2
\end{equation}
is defined as the \textit{variance of} $f_{\mu}$ and, as usually,
$\Delta f_{\mu}=\sqrt{(\Delta f_{\mu})^{2}}$ is the
\textit{uncertainty in} $f_{\mu}$.

Note that the inequality (\ref{227}) can be understood as the \textit{Schwarz inequality}.
Analogously (\ref{225}) leads to the \textit{Heisenberg -- Robertson uncertainty relation}
\begin{eqnarray}\label{229}
 (\Delta f_{1})^{2}\cdot (\Delta f_{2})^{2} &\geq& \left( {\rm Im} \left\{\int_{-\pi}^{\pi}(\delta\widehat{F}_{1}\psi(\phi))^{*} \delta\widehat{F}_{2}\psi(\phi) d\phi \right\}\right)^{2} \nonumber \\
   &= & \left( {\rm Im} \left\{\int_{-\pi}^{\pi} \psi^{*}(\phi)f_{1}^{*}(\phi)f_{2}\left(i\frac{\partial}{\partial\phi}\right)\psi(\phi) d\phi - \langle f_{1}(\phi)\rangle^{*}\langle f_{2}(n)\rangle \right\}
   \right)^{2}
\end{eqnarray}
Finally, the inequality (\ref{226}) gives the following \textit{Trifonov uncertainty relation}
\begin{equation}\label{230}
  (\Delta f_{1})^{2}+(\Delta f_{2})^{2} \geq 2 \left|{\rm Im} \left\{\int_{-\pi}^{\pi} \psi^{*}(\phi)f_{1}^{*}(\phi)f_{2}\left(i\frac{\partial}{\partial\phi}\right)\psi(\phi) d\phi -
  \langle f_{1}(\phi)\rangle^{*}\langle f_{2}(n)\rangle \right\}
  \right|.
\end{equation}

\begin{ex} \textit{Uncertainty relations for the phase and number of photons}
\label{ex2.1}
\end{ex}
Here we assume that $f_{1}=\phi$ and $f_{2}=n$. First, observe that we
should modify slightly the definition (\ref{228}) for $\Delta \phi$
to get a physically acceptable concept of the uncertainty in phase.
To this end we follow the results of D. Judge in his pioneering work
\cite{judge} and of H.S. Sharatchandra \cite{scharat}.

Consider then that $-\pi \leq \gamma < \pi$ and define $\delta \phi_{\gamma}:=\phi-\gamma$. Since $|\delta \phi_{\gamma}|$ can be greater than $\pi,$ we propose the following object
\setcounter{orange}{1}
\renewcommand{\theequation} {\arabic{section}.\arabic{equation}\theorange}
\begin{equation}\label{231a}
  \widetilde{\delta \phi_{\gamma}} = \left\{ \begin{array}{rl}
 \phi-\gamma+2\pi &\mbox{ for $-\pi\leq\phi\leq -\pi+\gamma$} \\
  \phi-\gamma &\mbox{ for $-\pi+\gamma\leq\phi<\pi$}
       \end{array} \right.
\end{equation}
\addtocounter{orange}{1}
\addtocounter{equation}{-1}
if $\gamma \geq 0$ and
\begin{equation}\label{231b}
  \widetilde{\delta \phi_{\gamma}} = \left\{ \begin{array}{rl}
 \phi-\gamma &\mbox{ for $-\pi\leq\phi\leq \pi+\gamma$} \\
  \phi-\gamma-2\pi &\mbox{ for $\pi+\gamma\leq\phi<\pi$}
       \end{array} \right.
\end{equation}
if $\gamma < 0$.
\renewcommand{\theequation} {\arabic{section}.\arabic{equation}}
Then the \textit{variance of} $\phi$ is defined by
\begin{equation}\label{232}
  (\widetilde{\Delta \phi})^{2}:=\min_{-\pi\leq\gamma<\pi}\int_{-\pi}^{\pi}(\widetilde{\delta \phi_{\gamma}} \psi(\phi))^{*}\widetilde{\delta
  \phi_{\gamma}}\psi(\phi)d\phi.
\end{equation}
Performing simple calculations and employing the periodicity of $\psi(\phi)$ i.e. $\psi(\phi\pm 2\pi)=\psi(\phi)$ one obtains
\begin{equation}\label{233}
  (\widetilde{\Delta \phi})^{2}=\min_{-\pi\leq\gamma<\pi}\int_{-\pi}^{\pi}\phi^{2}
  |\psi(\phi+\gamma)|^{2}d\phi.
\end{equation}
This result suggests that it is convenient to introduce a new wave
function
\begin{equation}\label{234}
  \widetilde{\psi}(\phi):=\psi(\phi+\gamma_{0}),
\end{equation}
where $\gamma_{0}$ minimizes (\ref{233}).

Consequently, (\ref{233}) reads now
\begin{equation}\label{235}
  (\widetilde{\Delta \phi})^{2}=\int_{-\pi}^{\pi}\phi^{2}|\widetilde{\psi}(\phi)|^{2}
  d\phi.
\end{equation}
Then
\begin{equation}\label{236}
  \langle \widetilde{n} \rangle :=\int_{-\pi}^{\pi} \widetilde{\psi}^{*}(\phi) i\frac{\partial}{\partial\phi} \widetilde{\psi}(\phi)d\phi 
                            = \int_{-\pi}^{\pi} \psi^{*}(\phi) i\frac{\partial}{\partial\phi} \psi(\phi)d\phi=\langle n \rangle.
\end{equation}
Since
$$
\left. \frac{\partial}{\partial\gamma}\int_{-\pi}^{\pi}\phi^{2}\big|\psi(\phi+\gamma)\big|^{2}
d\phi \right|_{\gamma=\gamma_{0}}=0,
$$
one quickly gets
\begin{eqnarray*}
  0 &=& \left. \int_{-\pi}^{\pi} \phi^{2} \frac{\partial}{\partial\gamma} \big|\psi(\phi+\gamma)\big|^{2} d\phi \right|_{\gamma=\gamma_{0}} = \int_{-\pi}^{\pi} \phi^{2} \frac{\partial}{\partial\phi} \big|\psi(\phi+\gamma_{0})\big|^{2} d\phi \nonumber \\
   &=& -2 \int_{-\pi}^{\pi} \phi \big|\psi(\phi+\gamma_{0})\big|^{2} d\phi = -2 \int_{-\pi}^{\pi} \phi \big|\widetilde{\psi}(\phi)\big|^{2}
   d\phi.
\end{eqnarray*}
Finally, one has
\begin{equation}\label{237}
  \widetilde{\langle \phi \rangle} := \int_{-\pi}^{\pi} \phi |\widetilde{\psi}(\phi)|^{2}
  d\phi=0.
\end{equation}
For the present case we redefine the $2\times2$ Hermitian matrix (\ref{221}) as
\begin{eqnarray}\label{238}
  \widetilde{F}_{11} &=& \int_{-\pi}^{\pi}\left(\phi\widetilde{\psi}(\phi)\right)^{*} \phi \widetilde{\psi}(\phi) d\phi = (\widetilde{\Delta\phi})^{2}, \nonumber \\
  \widetilde{F}_{12} &=& \int_{-\pi}^{\pi}\left(\phi\widetilde{\psi}(\phi)\right)^{*}\cdot\left(i\frac{\partial}{\partial\phi}-\langle n \rangle\right)\widetilde{\psi}(\phi) d\phi = \widetilde{F}_{21}^{*}, \nonumber \\
  \widetilde{F}_{22} &=& \int_{-\pi}^{\pi}\left[\left(i\frac{\partial}{\partial\phi}-\langle n \rangle\right)\widetilde{\psi}(\phi)\right]^{*} \left[\left(i\frac{\partial}{\partial\phi}-\langle n \rangle\right)\widetilde{\psi}(\phi)\right] d\phi = (\widetilde{\Delta n})^{2} = (\Delta
  n)^{2}.
\end{eqnarray}
Using (\ref{238}) one can rewrite the Robertson -- Schr\"{o}dinger uncertainty relation (\ref{227}) in the form
\begin{equation}\label{239}
  (\widetilde{\Delta \phi})^{2} \cdot (\Delta n)^{2} \geq \left|\int_{-\pi}^{\pi}(\phi\widetilde{\psi}(\phi))^{*} \left(i\frac{\partial}{\partial\phi}-\langle n \rangle\right)
  \widetilde{\psi}(\phi) d\phi \right|^{2} 
   = \left|i \int_{-\pi}^{\pi} \widetilde{\psi}^{*}(\phi) \phi \frac{\partial \widetilde{\psi}(\phi)}{\partial\phi} d\phi - \widetilde{\langle \phi\rangle} \langle n \rangle
   \right|^{2}.
\end{equation}
Then, integrating by parts
\[
 i \int_{-\pi}^{\pi} \widetilde{\psi}^{*}(\phi) \phi \frac{\partial \widetilde{\psi}(\phi)}{\partial\phi} d\phi = \frac{i}{2} \int_{-\pi}^{\pi} \left(\widetilde{\psi}^{*}(\phi) \phi \frac{\partial \widetilde{\psi}(\phi)}{\partial\phi}  + \widetilde{\psi}(\phi) \phi \frac{\partial \widetilde{\psi}^{*}(\phi)}{\partial\phi} \right) d\phi 
 \]
 \[
 + \frac{i}{2} \int_{-\pi}^{\pi} \left(\widetilde{\psi}^{*}(\phi) \phi \frac{\partial \widetilde{\psi}(\phi)}{\partial\phi} - \widetilde{\psi}(\phi) \phi \frac{\partial \widetilde{\psi}^{*}(\phi)}{\partial\phi} \right)d\phi
\]
\begin{equation}
\label{240}
= -\frac{i}{2} \bigg( 1 - 2\pi \big|\widetilde{\psi}(\pi)\big|^{2} \bigg) 
  + \frac{i}{2} \int_{-\pi}^{\pi} \left(\widetilde{\psi}^{*}(\phi) \phi \frac{\partial \widetilde{\psi}(\phi)}{\partial\phi}
   - \widetilde{\psi}(\phi) \phi \frac{\partial \widetilde{\psi}^{*}(\phi)}{\partial\phi}
   \right)d\phi.
\end{equation}
Substituting (\ref{240}) into (\ref{239}) and using (\ref{237}) we obtain the
Robertson -- Schr\"{o}dinger uncertainty relation for $\phi$ and $n$ in
the final form
\begin{equation}\label{241}
  (\widetilde{\Delta \phi})^{2} \cdot (\Delta n)^{2} \geq \frac{1}{4} \bigg( 1 - 2\pi \big|\widetilde{\psi}(\pi)\big|^{2} \bigg)^{2} + \left[ \frac{i}{2} \int_{-\pi}^{\pi} \left(\widetilde{\psi}^{*}(\phi) \phi \frac{\partial \widetilde{\psi}(\phi)}{\partial\phi}
   - \widetilde{\psi}(\phi) \phi \frac{\partial \widetilde{\psi}^{*}(\phi)}{\partial\phi}\right) d\phi \right]^{2}. 
\end{equation}

One immediately concludes that the Heisenberg -- Robertson uncertainty relation (\ref{229}) gives now
\begin{equation}\label{242}
  (\widetilde{\Delta \phi})^{2} \cdot (\Delta n)^{2} \geq \frac{1}{4} \bigg( 1 - 2\pi \big|\widetilde{\psi}(\pi)\big|^{2} \bigg)^{2}
\end{equation}
and the Trifonov relation (\ref{230}) reads
\begin{equation}\label{243}
  (\widetilde{\Delta \phi})^{2} + (\Delta n)^{2} \geq  \Big|1 - 2\pi |\widetilde{\psi}(\pi)|^{2}
  \Big|.
\end{equation}

\noindent
\textbf{Remark} The Heisenberg -- Robertson uncertainty
relation for the phase $\phi$ and the number of photons $n$ or for
the angle $\theta$ and the angular momentum $L_z$
were considered by many authors and  formulae analogous to
(\ref{242}) have been found
\cite{luks3,luks4,brif2,skagerstam,judge,pegg1}. The problem of
physically acceptable definition of the uncertainty in angle
$\theta$ or in phase $\phi$ was considered by D. Judge \cite{judge},
H.S. Sharatchandra \cite{scharat} or B-S. K. Skagerstam and B.A.
Bergsjordet \cite{skagerstam}. Our choice of the uncertainty
$\widetilde{\Delta \phi}$ (\ref{233}) is in accordance with those
works.

Now we are at the point, where the intelligent states for the phase
and the number of photons should be investigated. According to the
commonly used definition the \textit{Robertson -- Schr\"{o}dinger
intelligent state for $\phi$ and $n$} is a state represented by a function $\psi(\phi)$
such that the inequality (\ref{241}) reduces to the strict equality.
From the general results found in \cite{trifonov3,PT} or from a
careful analysis of the origin of the Schwarz inequality one
concludes that $\psi(\phi)$ is  an intelligent state if and
only if there exists $\lambda\in\mathbb{C}$ such that the following
equation
\begin{equation}\label{244}
  \left[ \left(i \frac{\partial}{\partial \phi} - \langle n \rangle \right) + i\lambda \phi \right]\widetilde{\psi}(\phi)=0
\end{equation}
is satisfied by a function $\widetilde{\psi}(\phi)$ related to the function $\psi(\phi)$  according to the rule (\ref{234}).

The general solution of Eq. (\ref{244}) reads
\begin{equation}\label{245}
\widetilde{\psi}(\phi)=\widetilde{a}e^{-i\langle n\rangle \phi}
e^{-\frac{\lambda}{2}\phi^{2}}, \,\, \ \ \  -\pi \leq \phi < \pi,
\end{equation}
where $\widetilde{a}\in \mathbb{C}$.

Now one should remember that some restrictions must be imposed on
the function $ \psi(\phi)$ if this function is going to
represent a photon state. First, the function $\psi(\phi)$ must be a periodic
function with period $2\pi$. Hence, $\widetilde{\psi}(\phi)$ is
also periodic with the same period $2\pi$ and, consequently
\begin{equation}\label{246}
\widetilde{\psi}(\pi)=\widetilde{\psi}(-\pi).
\end{equation}
From (\ref{245}) and (\ref{246}) one quickly gets the condition
\begin{equation}\label{247}
\langle n\rangle =N,\ \ \ \ \ \ N=0,1,2,...
\end{equation}
The second restriction follows immediately from (\ref{218}) with
(\ref{24}). Namely, writing the state $\vert \psi \rangle$ in the form
\begin{equation}\label{248}
\vert \psi \rangle= \sum^{\infty}_{n=0}c_n \vert n\rangle, \,\, \ \
\ \ c_n\in \mathbb{C}
\end{equation}
and using (\ref{218}) and (\ref{24}) one obtains
\begin{equation}\label{249}
\psi (\phi)=\langle \phi \vert \psi \rangle
=\frac{1}{\sqrt{2\pi}}\sum^{\infty}_{n=0}c_n e^{-in\phi}.
\end{equation}
This means that the Fourier expansion of the photon wave function
$\psi(\phi)$ must be of the form (\ref{249}) i.e. it does not
involve the exponents of the form $e^{in\phi},$ $n=1,2,...$

Analogously
\begin{equation}\label{250}
\widetilde{\psi}(\phi)=\psi(\phi +\gamma_0 )=\frac{1}{\sqrt{2\pi}}\sum^{\infty}_{n=0}c_ne^{-in(\phi +\gamma_0)}
=\frac{1}{\sqrt{2\pi}}\sum_{n=0}^{\infty}
\widetilde{c}_ne^{-in\phi}, \ \ \ \ \ \
\widetilde{c}_n=c_ne^{-in\gamma_0}.
\end{equation}
From (\ref{245}), (\ref{247}) and (\ref{250}) one infers that
\begin{equation}\label{251}
\int_{-\pi}^{\pi}e^{-\frac{\lambda}{2}\phi^{2}}e^{-i(N+k)\phi}d\phi=0
\ \ \ \ \  \textup{for} \ \ \  k=1,2,...
\end{equation}
and, consequently, also
\begin{equation}\label{252}
\int_{-\pi}^{\pi}e^{-\frac{\lambda}{2}\phi^{2}}e^{i(N+k)\phi}d\phi=0
\ \ \ \ \ \textup{for} \ \ \  k=1,2,...
\end{equation}
So the Fourier expansion of the function $e^{-\frac{\lambda}{2}\phi^{2}}$ satisfying (\ref{251}) and (\ref{252}) takes the form
\begin{equation}\label{253}
e^{-\frac{\lambda}{2}\phi^{2}}=\sum^{N}_{n=0}b_n \cos n\phi, \ \ \ \
\ \ b_n \in \mathbb{C}.
\end{equation}
Differentiating both sides of (\ref{253}) with respect to $\phi$ and putting $\phi=-\pi$ we get
\begin{equation}\label{254}
\lambda \pi e^{-\frac{\lambda}{2}\pi^{2}}= \sum^{N}_{n=1}nb_n \sin
n\pi=0.
\end{equation}
Eq. (\ref{254}) holds true if and only if $\lambda=0$. Hence, by
(\ref{245}) and (\ref{234}) one finds $\psi(\phi)$ normalized to $1$
as
\begin{equation}\label{255}
\psi(\phi)=\frac{1}{\sqrt{2\pi}}e^{-iN\phi}, \ \ \ \ \ \  N=0,1,...
\end{equation}
(Note that now we can put $ \gamma_0 =0$). So the respective ket  $\vert \psi \rangle$ reads
\begin{equation}\label{256}
\vert \psi \rangle=\vert N \rangle, \ \ \ \ \ \ N=0,1,...
\end{equation}
One quickly shows that for any of the states (\ref{256}) the relations are satisfied
\setcounter{orange}{1}
\renewcommand{\theequation} {\arabic{section}.\arabic{equation}\theorange}
\begin{equation}\label{257a}
\Delta n = 0,
\end{equation}
\addtocounter{orange}{1}
\addtocounter{equation}{-1}
\begin{equation}\label{257b}
 \widetilde{\Delta \phi}=\Delta \phi = \frac{\pi}{\sqrt{3}}
\end{equation}
\renewcommand{\theequation} {\arabic{section}.\arabic{equation}}

and also that the states (\ref{256})  are the \textit{Heisenberg -- Robertson
intelligent states} for $\phi$ and $n$ i.e. they fulfill the strict
equality in (\ref{242}) but by (\ref{257a}) and (\ref{257b}) none of
them gives the strict equality for (\ref{243}), so none of them can
be a \textit{Trifonov intelligent state} for $\phi$ and $n$. Then,
since the following implication 
$\textrm{Eq.}$ (\ref{226})
$\Rightarrow \textrm{Eq.}$ (\ref{225}) $\Rightarrow
\textrm{Eq.}$ (\ref{224}) 
holds true (see \cite{PT}), one easily
concludes that any number -- phase Trifonov intelligent state is a
number -- phase Heisenberg -- Robertson intelligent state and also a
number -- phase Robertson -- Schr\"{o}dinger intelligent state. Finally,
gathering all that we arrive at the result:

\begin{tw}
\label{t2.1}
{\rm
The only number -- phase Robertson -- Schr\"{o}dinger intelligent states
are the eigenstates of the number operator $\widehat{n}$, $\vert
n\rangle, n=0,1,...$ These states are also the only number -- phase
Heisenberg -- Robertson intelligent states. There are no number -- phase Trifonov
intelligent states.} \rule{2mm}{2mm}
\end{tw}

\section{Intelligent States for Arbitrary $f_1$ and $f_2$}
\setcounter{ex}{0}
\setcounter{tw}{0}
\setcounter{co}{0}

\setcounter{equation}{0}

From the results of \cite{trifonov1,trifonov2,trifonov3,trifonov4,trifonov5,dodonov,PT} we conclude that the \textit{Robertson -- Schr\"{o}dinger intelligent state for $f_1=f_1(\phi)$ and $f_2=f_2(n)$} can be found as a solution of the following equation
\begin{equation}\label{31}
\left( \delta \widehat{F}_2 +i\lambda \delta \widehat{F}_1 \right) \psi(\phi)=0,
\end{equation}
where $\lambda \in \mathbb{C}$, $\delta \widehat{F}_{1}$ and $\delta \widehat{F}_{2}$ are defined by (\ref{220a}) and (\ref{220b}) with (\ref{219a}) and (\ref{219b}). Eq. (\ref{31}) can be rewritten in the form
\begin{equation}\label{32}
\left[ f_2 \left(i\frac{\partial}{\partial \phi}\right)+i\lambda
f_1(\phi)-\mu \right] \psi(\phi)=0, \ \ \ \ \  \lambda \in
\mathbb{C}
\end{equation}
with
\begin{equation}\label{33}
\mu=\langle f_2(n)\rangle  +i\lambda\langle f_1(\phi)\rangle .
\end{equation}
Moreover, as in the previous case when $f_1=\phi$ and $f_2=n$, on the state $\psi(\phi)$ the conditions
\begin{equation}\label{34}
\psi(\pi)=\psi(-\pi)
\end{equation}
and (\ref{249}) are
imposed. One quickly gets that Eq. (\ref{32}) restricted to $\lambda \in {\mathbb R} $ defines the Heisenberg -- Robertson intelligent states.

\noindent
\textbf{Remark} Our equation (\ref{32}) is different from
the respective equation which one could find by using the considerations
analogous to those given by C. Brif and Y. Ben -- Aryeh \cite{brif1}.
The reason lies in the fact that, in general
\begin{equation}\label{35}
\langle \phi \vert \widehat{f}_1(\phi)\vert \psi \rangle \neq f_1(\phi)\langle \phi \vert \psi \rangle =f_1(\phi) \cdot \psi(\phi),
\end{equation}
where $\widehat{f}_1(\phi)$ is defined by (\ref{27}).
Note that  contrary to (\ref{35}) one gets
\begin{equation}\label{36}
\langle \phi \vert \widehat{f}_2(n)\vert \psi \rangle=f_2
\left(i\frac{\partial}{\partial \phi}\right)\langle \phi \vert \psi
\rangle=f_2 \left(i\frac{\partial}{\partial \phi}\right)\psi (\phi).
\end{equation}

As the first example of Eq. (\ref{32}) let us  consider the following case

\begin{ex}
\label{ex3.1}
\end{ex}
\begin{equation}\label{37}
f_1(\phi)=e^{i\phi},\ \ \ \ f_2(n)=n.
\end{equation}
Eq. (\ref{32}) reads now
\begin{equation}\label{38}
\left( i\frac{\partial}{\partial \phi}+i\lambda e^{i\phi}-\mu
\right) \psi(\phi)=0,\ \ \ \   \lambda \in \mathbb{C}.
\end{equation}
The general solution of (\ref{38}) is
\begin{equation}\label{39}
\psi(\phi)=ae^{-i\mu \phi}\cdot e^{i\lambda e^{i\phi}}, \ \ \ a\in \mathbb{C}.
\end{equation}
The condition (\ref{34}) yields $\mu \in \mathbb{Z}$. Expanding the term $e^{i \lambda e^{i\phi}}$ in (\ref{39}) one has
\begin{equation}\label{310}
\psi(\phi)=ae^{-i\mu \phi} \sum_{k=0}^{\infty}\frac{(i\lambda)^k}{k!}e^{ik\phi}.
\end{equation}
Then $\psi(\phi)$ given by (\ref{310}) fulfills the condition (\ref{249}) iff
\begin{equation}\label{311}
\lambda=0,\ \ \ \mu=n,\ \ \ n=0,1,...
\end{equation}
So the only normalized states $\psi(\phi)$ satisfying Eq. (\ref{38}) are again the eigenstates of $\widehat{n}$, $\psi(\phi)=\frac{1}{\sqrt{2\pi}}e^{-in\phi}$ with $n=0,1,...$ Gathering, analogously as in Example \ref{ex2.1} \textit{the only Robertson -- Schr\"{o}dinger and Heisenberg -- Robertson intelligent states for $e^{i\phi}$ and $n$ are the eigenstates of $\widehat{n}$, $\vert n\rangle$, $n=0,1,...$. There are no Trifonov intelligent states for $e^{i\phi}$ and $n$}.

\begin{ex} \label{ex3.2}
\end{ex}

Here we assume
\begin{equation}\label{312}
f_1(\phi)=e^{-i\phi},\ \ \ f_2(n)=n.
\end{equation}
Substituting (\ref{312}) into Eq. (\ref{32}) we have
\begin{equation}\label{313}
\left( i\frac{\partial}{\partial \phi}+i\lambda e^{-i\phi}-\mu \right)\psi (\phi)=0, \ \ \ \lambda \in \mathbb{C}.
\end{equation}
The general normalized solution of (\ref{313}) satisfying the conditions (\ref{34}) and (\ref{249}) reads
\begin{equation}\label{314}
\psi(\phi)={\left( 2\pi I_0 (2\vert \lambda\vert)\right)}^{-\frac{1}{2}} e^{-in\phi} e^{-i\lambda e^{-i\phi}}
= {\left( 2\pi I_0 (2\vert \lambda\vert)\right)}^{-\frac{1}{2}} \sum_{k=0}^{\infty}\frac{(-i\lambda)^k}{k!}e^{-i(n+k)\phi},
\end{equation}
\[
\mu = n=0,1,..., \ \ \ \lambda \in \mathbb{C}, 
\]
where $I_0$ is the zeroth modified Bessel function of the first kind. In the Dirac notation the state $\vert \psi\rangle$ is of the form
\begin{equation}\label{315}
\vert \psi\rangle={\big( I_0(2\vert \lambda\vert)\big)}^{-\frac{1}{2}}\sum_{k=0}^{\infty}\frac{(-i\lambda)^k}{k!} \vert n+k\rangle, \ \ \lambda \in \mathbb{C}, \ \ n=0,1,...
\end{equation}
\textbf{Remark} The solution (\ref{314}) was found by C. Brif and Y. Ben -- Aryeh \cite{brif1,brif2} and then, with the use of Hardy space formalism, by S. Luo \cite{luo}. Note that in the present case one has that
\begin{equation}\label{316}
\widehat{e^{-i\phi}}=\int_{-\pi}^{\pi} e^{-i\phi}\vert \phi\rangle \langle \phi \vert \ d\phi= \sum_{n=0}^{\infty}\vert n+1\rangle \langle n\vert
\end{equation}
and this is an exceptional case when,  contrary to (\ref{35}), we get the equality
\begin{equation}\label{317}
\langle \phi \vert \widehat{e^{-i\phi} }\vert \psi \rangle =e^{-i\phi} \langle \phi\vert \psi\rangle= e^{-i\phi} \psi(\phi).
\end{equation}
So our equation (\ref{313}) is equivalent to Eq. (4.29) of Ref. \cite{brif1} and, consequently, our solution (\ref{314}) is the same as the respective solution (4.31) of \cite{brif1}.

Straightforward calculations give
\setcounter{orange}{1}
\renewcommand{\theequation} {\arabic{section}.\arabic{equation}\theorange}
\begin{equation}\label{318a}
\langle n\rangle = n+\vert \lambda \vert \frac{I_1(2\vert \lambda \vert)}{I_0(2\vert \lambda \vert)},
\end{equation}
\addtocounter{orange}{1}
\addtocounter{equation}{-1}
\begin{equation}\label{318b}
  \langle e^{-i\phi}\rangle = i\frac{\lambda^* }{\vert \lambda \vert}\frac{I_1(2\vert \lambda \vert)}{I_0(2\vert \lambda \vert)},
\end{equation}
\addtocounter{orange}{1}
\addtocounter{equation}{-1}
\begin{equation}\label{318c}
  \langle n^2 \rangle = n^2+ {\vert \lambda \vert}^2+2n \vert \lambda \vert \frac{I_1(2\vert \lambda \vert)}{I_0(2\vert \lambda \vert)},
\end{equation}
\addtocounter{orange}{1}
\addtocounter{equation}{-1}
\begin{equation}\label{318d}
  (\Delta n)^2 = {\vert \lambda \vert}^2 \left[ 1- {\left( \frac{I_1(2\vert \lambda \vert)}{I_0(2\vert \lambda \vert)}\right)^2} \right],
\end{equation}
\addtocounter{orange}{1}
\addtocounter{equation}{-1}
\begin{equation}\label{318e}
  (\Delta e^{-i\phi})^2 = 1- \left( \frac{I_1(2\vert \lambda \vert)}{I_0(2\vert \lambda \vert)} \right)^2,
\end{equation}
\addtocounter{orange}{1}
\addtocounter{equation}{-1}
\begin{equation}\label{318f}
{\left( {\rm Im} \left\lbrace
\int_{-\pi}^{\pi}\psi^{*}(\phi)(e^{-i\phi})^*
i\frac{\partial}{\partial \phi} \psi(\phi)d\phi - {\langle
e^{-i\phi}\rangle}^*\langle n\rangle\right\rbrace \right)}^2= ({\rm
Re} \lambda)^2 {\left[ 1-{\left( \frac{I_1(2\vert \lambda
\vert)}{I_0(2\vert \lambda \vert)}\right)}^2 -\frac{n}{|\lambda|}   \frac{I_1(2\vert \lambda
\vert)}{I_0(2\vert \lambda \vert)} \right] }^2
\end{equation}
\renewcommand{\theequation} {\arabic{section}.\arabic{equation}}
(compare with respective formulae of \cite{brif1,brif2}).

Using the above relations one quickly arrives at the following result.

\begin{tw} \label{t3.1}{\rm

The Robertson -- Schr\"{o}dinger intelligent states for $e^{-i\phi}$
and $n$ are given by the wave functions (\ref{314}) (or
equivalently, by the vectors (\ref{315})). The Heisenberg -- Robertson
intelligent states for $e^{-i\phi}$ and $n$ are given by the
functions (\ref{314}) ($\equiv$ the vectors (\ref{315})) with
$\lambda \in \mathbb{R}$. Finally, the Trifonov intelligent states
for $e^{-i\phi}$ and $n$ are given by the functions
(\ref{314}) ($\equiv$ the vectors (\ref{315})) with $\lambda = \pm
1$. }
\rule{2mm}{2mm}
\end{tw}

From (\ref{318d}) and (\ref{318e}) it follows that
\begin{equation}\label{319}
\Delta n = \Delta e^{-i\phi} \ \ \  \Leftrightarrow \ \ \ \vert \lambda \vert=1.
\end{equation}
For completeness, we give now the uncertainties $\Delta \cos \phi$ and $\Delta \sin \phi$ in the state (\ref{314}).

Simple calculations show that (see \cite{brif1})
\begin{equation}\label{320}
{\left( \Delta \cos \phi\right)}^2=\langle \cos^2
\phi\rangle-{\langle \cos \phi \rangle}^2 = \frac{1}{2}+\frac{({\rm
Im} \lambda)^2-({\rm Re} \lambda)^2}{2\vert \lambda\vert^{2}}
\frac{I_2(2\vert \lambda\vert)}{I_0 (2\vert \lambda\vert)}-
\frac{({\rm Im} \lambda)^2}{{\vert \lambda\vert}^2} {\left(
\frac{I_1(2\vert \lambda\vert)}{I_0 (2\vert \lambda\vert)}\right)}^2
\end{equation}
and
\begin{equation}\label{321}
{\left( \Delta \sin \phi\right)}^2=\langle \sin^2
\phi\rangle-{\langle \sin \phi \rangle}^2 = \frac{1}{2}+\frac{({\rm
Re} \lambda)^2-({\rm Im} \lambda)^2}{2\vert \lambda\vert^{2}}
\frac{I_2(2\vert \lambda\vert)}{I_0 (2\vert \lambda\vert)}-
\frac{({\rm Re} \lambda)^2}{{\vert \lambda\vert}^2} {\left(
\frac{I_1(2\vert \lambda\vert)}{I_0 (2\vert
\lambda\vert)}\right)}^2,
\end{equation}
where $I_1$ and $I_2$ are the first and the second modified Bessel functions of the first kind, respectively.

Adding (\ref{320}) and (\ref{321}), and comparing with (\ref{318e}) one gets
\begin{equation}\label{322}
{\left( \Delta \cos \phi\right)}^2+{\left( \Delta \sin \phi\right)}^2=(\Delta e^{-i\phi})^2=1-{\left( \frac{I_1(2\vert \lambda\vert)}{I_0 (2\vert \lambda\vert)}\right)}^2.
\end{equation}
In particular, from (\ref{318d}), (\ref{320}) and (\ref{321}) for $\lambda=\pm 1$ we have
\setcounter{orange}{1}
\renewcommand{\theequation} {\arabic{section}.\arabic{equation}\theorange}
\begin{equation}
\label{323a}
(\Delta \cos \phi)^2= \frac{1}{2} -  \frac{1}{2} \frac{I_2(2)}{I_0(2)} \approx 0.349,
\end{equation}
\addtocounter{orange}{1}
\addtocounter{equation}{-1}
\begin{equation}
\label{323b}
(\Delta \sin \phi)^2= \frac{1}{2}  +  \frac{1}{2} \frac{I_2(2)}{I_0(2)}  -{\left( \frac{I_1(2)}{I_0(2)}\right)}^2=(\Delta n)^2-\frac{1}{2}+  \frac{1}{2} \frac{I_2(2)}{I_0(2)} \approx 0.164,
\end{equation}
\addtocounter{orange}{1}
\addtocounter{equation}{-1}
\begin{equation}
\label{323c}
(\Delta \cos \phi)^2+(\Delta \sin \phi)^2=1-{\left(
\frac{I_1(2)}{I_0(2)}\right)}^2 \approx 0.513.
\end{equation}
\renewcommand{\theequation} {\arabic{section}.\arabic{equation}}  

\begin{ex} \label{ex3.3}
\end{ex}
We consider here the following case
\begin{equation}\label{324}
f_1(\phi)=\cos \phi, \ \ \ f_2(n)=n.
\end{equation}
Then Eq. (\ref{32}) reads now
\begin{equation}\label{325}
\left( i\frac{\partial}{\partial \phi} +i\lambda \cos \phi - \mu
\right) \psi(\phi)=0, \ \ \ \ \ \ \lambda \in \mathbb{C}.
\end{equation}
The normalized to $1$ and satisfying the periodicity condition (\ref{34}) solution of (\ref{325})  is
\begin{equation}\label{326}
\psi(\phi)=\frac{1}{\sqrt{2\pi I_0 (2{\rm
Re}\lambda)}}e^{-in\phi}e^{-\lambda \sin \phi}, \ \ \ \ \
\mu=n=0,1,...
\end{equation}
Note that ${\vert \psi (\phi)\vert}^2= \frac{1}{2\pi I_0 (2{\rm
Re}\lambda)}e^{-2{\rm Re}\lambda \cdot \sin \phi}$ is the von Mises
circular distribution.

Then, the solution (\ref{326}) fulfills the condition (\ref{249}) iff
\begin{equation}\label{327}
\int_{-\pi}^{\pi}e^{-\lambda \sin \phi} e^{-i(n+k)\phi}d\phi=0, \ \
\ \ \ \textup{for} \ \ k=1,2,...
\end{equation}
As is well known
\begin{equation}\label{328}
\int_{-\pi}^{\pi}e^{-\lambda \sin \phi}e^{-im\phi}d\phi=2\pi
J_m(i\lambda), \ \ \ \ \  m=0,1,...
\end{equation}
where $J_m$ denotes the $m$-th Bessel function of the first kind.
From (\ref{327}) and (\ref{328}) one infers that $\psi(\phi)$ given
by (\ref{326}) satisfies the condition (\ref{249}) iff
\begin{equation}\label{329}
J_{n+k}(i\lambda)=0, \ \  \ \ \ \textup{for\; all} \ \ k=1,2,...
\end{equation}
But the system of constraints (\ref{329}) holds true iff $\lambda=0$.

So \textit{the only Robertson -- Schr\"{o}ndinger and
Heisenberg -- Robertson intelligent states for $\cos \phi$ and $n$ are
given by the normalized wave functions}
$\psi(\phi)=\frac{1}{\sqrt{2\pi}}e^{-in\phi}$, $n=0,1,...$,
\textit{i.e. the eigenfunctions of the number operator
$\widehat{n}$}.

Then one can easily find that the uncertainty $\Delta \cos \phi$ in
the state $\psi(\phi)=\frac{1}{\sqrt{2\pi}}e^{-in\phi}$,
$n=0,1,...$, is given by
\begin{equation}\label{330}
\Delta \cos \phi= \frac{1}{\sqrt{2}}
\end{equation}
and, consequently
\begin{equation}\label{331}
(\Delta n)^2 + (\Delta \cos \phi)^2=\frac{1}{2}.
\end{equation}
As can be quickly shown, the right hand side of the Trifonov
uncertainty relation (\ref{230}) in our present case vanishes. Thus the
formula (\ref{331}) shows immediately that the states
${\psi}(\phi)=\frac{1}{\sqrt{2\pi}}e^{-in\phi}$, $n=0,1,...$ are
not the Trifonov intelligent states and, therefore, \textit{there
are no Trifonov intelligent states for $\cos \phi$ and the photon
number $n$}.

Finally, we observe that some conclusions on intelligent states hold true when
\begin{equation}\label{332}
f_1=\sin \phi, \ \ \ f_2=n.
\end{equation}


\section{The Minimum Uncertainty Product States}
\setcounter{ex}{0}
\setcounter{tw}{0}
\setcounter{co}{0}

\setcounter{equation}{0}

In this section we deal with the problem of searching for the states
which {\it minimize locally} the product of uncertainties $(\Delta
f_1)^2 \cdot (\Delta f_2)^2$. To this end we consider the functional

\begin{equation}
{\cal H}[\psi(\phi),\psi^*(\phi)] :=  (\Delta f_1)^2 \cdot (\Delta f_2)^2.
\label{hamilton41}
\end{equation}
 Thus our problem reduces to the following isoperimetric
variational problem
\begin{equation}
\delta {\cal H}[\psi(\phi),\psi^*(\phi)] =0, \ \ \ \ \ \ \int_{- \pi}^{\pi}
\psi^*(\phi) \psi(\phi) d \phi =1.
\label{variation42}
\end{equation}
Inserting Eq. (\ref{hamilton41}) into Eq. (\ref{variation42}) and
employing Def. (\ref{228}) one easily  proves that the
variational problem (\ref{variation42}) leads to the  following
equation
\begin{equation}
\bigg[ (\Delta f_1)^2 \cdot (\delta \widehat{F}_2)^\dag \cdot \delta
\widehat{F}_2 +  (\Delta f_2)^2 \cdot (\delta \widehat{F}_1)^\dag
\cdot \delta \widehat{F}_1 - \sigma \bigg] \psi(\phi) =0,
\label{condition143}
\end{equation}
where $\sigma \in \mathbb{R}$ is a Lagrange multiplier. The operators $\delta
\widehat{F}_1$ and $\delta \widehat{F}_2$ are defined by  Eqs.
(\ref{220a}) and (\ref{220b}), respectively.

Multiplying (\ref{condition143}) by $\psi^*(\phi)$ and integrating  we find that
\begin{equation}
\sigma = 2 (\Delta f_1)^2 \cdot (\Delta f_2)^2.
\label{sigmass44}
\end{equation}
Substituting (\ref{sigmass44}) into (\ref{condition143}) and
dividing by $(\Delta f_1)^2 \not=0$ one finally gets
\begin{equation}
\bigg[(\delta \widehat{F}_2)^\dag \cdot \delta \widehat{F}_2 +
{(\Delta f_2)^2  \over (\Delta f_1)^2} (\delta \widehat{F}_1)^\dag
\cdot \delta \widehat{F}_1 - 2 (\Delta f_2)^2 \bigg] \psi(\phi) =0.
\label{generaleqn45}
\end{equation}
Eq. (\ref{generaleqn45}) for Hermitian operators
$\widehat{F}_1^\dag = \widehat{F}_1$ and $\widehat{F}_2^\dag =
\widehat{F}_2$ was  previously found by R. Jackiw \cite{jackiw}
and then also analyzed extensively by P. Carruthers and M.M. Nieto
\cite{carruthers2}. Our Eq. (\ref{generaleqn45}) is an obvious
generalization of their results on the case of non -- hermitian
operators.

Now we consider the first important example of the application
of Eq. (\ref{generaleqn45}).

\begin{ex} \label{ex4.1}
\end{ex}

Here we assume that $f_1$ and $f_2$ are given by (\ref{312}) i.e.
$f_1(\phi)= e^{-i\phi}$ and $f_2(n) = n$. Then Eq.
(\ref{generaleqn45}) reads now
\begin{equation}
\bigg\{\bigg( i {\partial \over \partial \phi} - \langle n
\rangle\bigg)^2 + {(\Delta n)^2 \over (\Delta
e^{-i\phi})^2}\bigg(e^{i \phi} - \langle e^{i \phi} \rangle \bigg)
\bigg(e^{-i \phi} - \langle e^{-i \phi} \rangle \bigg) - 2 (\Delta
n)^2 \bigg\} \psi(\phi) =0. \label{gendiffeqn46}
\end{equation}
One should remember that the wave function $ \psi(\phi)$ must
be of the form given by (\ref{249}). Substituting (\ref{249}) into
(\ref{gendiffeqn46}) we quickly note that the three following cases
should be analyzed:

\renewcommand{\theenumi}{(\roman{enumi})}
\begin{enumerate}
\item
\label{i147}
\begin{equation}
\Delta n \not=0, \ \ \ \ \ \big\langle e^{-i \phi} \big\rangle
\not=0. \label{cond147}
\end{equation}
Here one immediately finds that all coefficients $c_n$ in
(\ref{249}) vanish, so $\psi(\phi) =0$.
\item
\label{i248}
\begin{equation}
\Delta n \not=0, \ \ \ \ \ \ \  \big\langle e^{-i \phi} \big\rangle
=0.
\label{cond248}
\end{equation}
Then
\begin{equation}
\big( \Delta e^{-i \phi} \big)^2 = \int_{-\pi}^\pi \psi^*(\phi)
\bigg( e^{i \phi} - \langle e^{i \phi} \rangle \bigg) \bigg( e^{-i
\phi} - \langle e^{-i \phi} \rangle \bigg)\psi(\phi) d \phi =1
\label{deltaexp49}
\end{equation}
and Eq. (\ref{gendiffeqn46}) reduces to
\begin{equation}
\bigg\{\bigg( i {\partial \over \partial \phi} - \langle n \rangle
\bigg)^2 - (\Delta n)^2 \bigg\} \psi(\phi) =0.
\label{DiffEqn410}
\end{equation}
The general normalized to 1 solution of (\ref{DiffEqn410})
satisfying (\ref{cond248}) and  (\ref{249})
reads
\begin{equation}
\psi(\phi)= {1 \over \sqrt{4 \pi}} e^{i \alpha} \bigg( e^{-ik\phi} +
e^{i\beta}e^{-i\ell\phi} \bigg),
\label{solutionpsi411}
\end{equation}
where $\alpha, \beta \in \mathbb{R}$, $ {\mathbb N} \ni k, \ell \geq 0$, $|\ell -k|
\geq 2$. Straightforward calculations give
\begin{equation}
 \langle n \rangle ={k + \ell \over 2}, \ \ \ \ \ \Delta n = {|\ell - k| \over 2}.
\label{desigualdad1412}
\end{equation}

We will  show now that although the wave function (\ref{solutionpsi411})
fulfills Eq. (\ref{gendiffeqn46}), the respective quantum state is
not a minimum uncertainty product state. To this end assume that in
(\ref{solutionpsi411}) the natural numbers $k$ and $\ell$ are chosen
so that $k< \ell$. Define the normalized function
\begin{equation}
 \psi ' (\phi):= \sqrt{1 -\varepsilon}\psi(\phi) +
\sqrt{\varepsilon}{1 \over \sqrt{2 \pi}} e^{-im\phi},
\label{solutionpsiprime413}
\end{equation}
where $m \in \mathbb{N}$, $k<m<\ell,\, \varepsilon \in
\mathbb{R}$, $0<\varepsilon \leq 1,$ and $\psi(\phi)$ is given by
(\ref{solutionpsi411}).

Denoting the uncertainties of the photon number $n$ and the phase
function $e^{-i \phi}$ for the state (\ref{solutionpsiprime413}) by
$(\Delta n)'$ and $(\Delta e^{-i\phi})'$, respectively one easily
gets
$$
(\Delta n)'^{\,2} =\bigg({k-\ell \over 2}\bigg)^2 + \varepsilon
(k-m)(\ell -m) - \varepsilon^2 \bigg( {k+\ell \over 2} -m\bigg)^2
$$
\setcounter{orange}{1}
\renewcommand{\theequation} {\arabic{section}.\arabic{equation}\theorange}
\begin{equation}
= (\Delta n)^{\,2} +  \varepsilon (k-m)(\ell -m) - \varepsilon^2\big(
\langle n \rangle - m \big)^2 < (\Delta n)^2,
\label{deltan414a}
\end{equation}
\addtocounter{orange}{1}
\addtocounter{equation}{-1}
\begin{equation}
\big(\Delta e^{-i\phi}\big)'^2 =1 - \big| \langle e^{-i\phi} \rangle
'\big|^2 \leq 1 = \big(\Delta e^{-i\phi}\big)^2,
\label{deltae414b}
\end{equation}
\renewcommand{\theequation} {\arabic{section}.\arabic{equation}}  
where $\big(\Delta e^{-i\phi}\big)^2$, $\Delta n$ and $\langle n
\rangle$ are  given by (\ref{deltaexp49}) and
(\ref{desigualdad1412}), and $\big\langle e^{-i \phi} \big\rangle'
:= \int_{-\pi}^{\pi} e^{-i \phi} |\psi'(\phi)|^2 d \phi$.

The norm of the difference $\psi' - \psi$ is given by
$$
\| \psi' - \psi \|_0 := \sup_{-\pi \leq \phi < \pi} | \psi'(\phi) -
\psi(\phi)| \leq \big| \sqrt{1 - \varepsilon} -1 \big| \| \psi \|_0
+ \sqrt{\varepsilon \over 2 \pi}
$$
\begin{equation}
\leq {1 - \sqrt{1 - \varepsilon} \over \sqrt{\pi}} +
\sqrt{\varepsilon \over 2 \pi} = \sqrt{\varepsilon \over 2 \pi}
\bigg({\sqrt{2\varepsilon} \over 1 + \sqrt{1 - \varepsilon}} + 1
\bigg), \ \ \ \ \ \ \ 0 < \varepsilon \leq 1 \label{diferencia415}
\end{equation}
and it can be done  arbitrarily small by taking $\varepsilon$
sufficiently small. Since by (\ref{deltan414a}) and
(\ref{deltae414b})
\begin{equation}
(\Delta n)' \cdot (\Delta e^{-i \phi})' < \Delta n \cdot  \Delta
e^{-i \phi}, \ \ \ \ \ \  \forall \ \ 0 < \varepsilon \leq 1,
\label{producto416}
\end{equation}
the state (\ref{solutionpsi411}) is {\it not a minimum uncertainty
product state for $n$ and $e^{-i \phi}$}. Note that taking in
(\ref{solutionpsiprime413})
\begin{equation}
m > \ell + 1 > k + 1
\label{desigualdad2417}
\end{equation}
and putting $\varepsilon$ sufficiently small one easily infers from
 Eqs. (\ref{deltan414a}) and (\ref{deltae414b}) that, with
(\ref{desigualdad2417}) assumed, the relations 
$$
(\Delta n)' > \Delta n, \ \ \ \ \ \ \ \ (\Delta e^{-i \phi})' =1
$$
imply
\begin{equation}
(\Delta n)' \cdot (\Delta e^{-i \phi})' > \Delta n \cdot  \Delta
e^{-i \phi}.
\label{desigprime418}
\end{equation}
Therefore, the state (\ref{solutionpsi411}) {\it is also not a
maximum uncertainty product state}. Remember that we deal with a  {\it
local} minimum and a {\it  local} maximum uncertainty product states.

It remains the last case to be considered
\item
\label{cond419}
\begin{equation}
\Delta n =0.
\label{419}
\end{equation}
Here of course, $\psi(\phi)= {1 \over \sqrt{2 \pi}} e^{-in \phi}$,
$n=0,1,2, \dots,$  $\langle n \rangle =n,$ $\big\langle e^{-i\phi}
\big\rangle =0,$ $\Delta e^{-i \phi} =1$ and $\Delta n \cdot \Delta
e^{-i \phi} =0$.
\end{enumerate}

 Thus we get now the (global) minimum uncertainty product
states. Summarizing, one arrives at 

\begin{tw}
\label{t4.1}
{\rm

The only minimum uncertainty product states for the photon number
$n$ and the phase function $e^{-i \phi}$ are the eigenstates $|n\rangle$,
$n=0,1,2, \dots $ of  the photon number operator $\widehat{n}$.

The states $|n\rangle$ are the global minimum uncertainty product states
for $n$ and $e^{-i \phi}$.} \rule{2mm}{2mm}
\end{tw}
From this theorem one immediately gets

\begin{co}
\label{co4.1}
{\rm

If $|\psi \rangle$, $\langle \psi | \psi \rangle =1$, is a state
different from the eigenstate of $\widehat{n}$, and $\Delta n$ and
$\Delta e^{-i \phi}$ are uncertainties for the number of photons and
for the phase function $e^{-i\phi}$ respectively, then in any
neighborhood of   $|\psi \rangle$ (in the sense of the norm $\|
\cdot \|_0$) there exists a state  $|\psi' \rangle$, $\langle \psi'
| \psi' \rangle =1$, such that the product of uncertainties $(\Delta
n)'$ and $(\Delta e^{-i \phi})'$ is less than the product of $\Delta
n$ and $\Delta e^{-i \phi}$ i.e. $(\Delta n)' \cdot (\Delta e^{-i
\phi})' < \Delta n \cdot \Delta e^{-i \phi}$.}  \rule{2mm}{2mm}
\end{co}

One can quickly observe that Theorem \ref{t4.1} and Corollary \ref{co4.1} are also
true, {\it mutatis mutandi}, for the cases $f_1=e^{i\phi}$, $f_1
=\cos \phi$ or $f_1= \sin \phi$ and $f_2 =n$.

The second important example we deal with is

\begin{ex}
\label{ex4.2}
\end{ex}
Here we look for the minimum uncertainty product states for the
quantum phase and the number of photons.

A careful analysis of Example \ref{ex2.1} shows that one should consider now
the following isoperimetric variational problem
\begin{equation}
\delta \bigg\{ \int_{-\pi}^{\pi} \phi^2 |\psi(\phi)|^2 d \phi \cdot
\int_{- \pi}^\pi \psi^*(\phi) \bigg(i{\partial \over \partial \phi}
- \langle n \rangle \bigg)^2 \psi(\phi) d \phi \bigg\} =0,
\end{equation}
with the constraint  $\int_{-\pi}^\pi \psi^*(\phi) \psi(\phi) d \phi =1.$

 Then in the present case the counterpart of Eq.
(\ref{generaleqn45}) reads
\begin{equation}
\bigg\{\bigg( i {\partial \over \partial \phi} - \langle n
\rangle\bigg)^2 + {(\Delta n)^2 \over \langle \phi^2 \rangle} \phi^2
- 2 (\Delta n)^2 \bigg\} \psi(\phi) =0.
\label{421}
\end{equation}
Substituting
\begin{equation}
 \chi(\phi) = e^{i \langle n \rangle\phi} \psi(\phi)
\label{422}
\end{equation}
we turn Eq. (\ref{421}) to the form
\begin{equation}
{d^2 \chi(\phi) \over d \phi^2} - \bigg[ {(\Delta n)^2 \over \langle
\phi^2 \rangle} \phi^2 - 2 (\Delta n)^2 \bigg]  \chi(\phi) =0.
\label{423}
\end{equation}
Assume first that $\Delta n \not=0$ and define a new variable
\begin{equation}
z = \bigg({2 \Delta n \over \sqrt{\langle \phi^2 \rangle}}
\bigg)^{1/2} \phi. \label{424}
\end{equation}
Then Eq. (\ref{423}) reads
\begin{equation}
{d^2 \chi(z) \over d z^2} - \bigg( {z^2 \over 4} - \Delta n \cdot
\sqrt{\langle \phi^2 \rangle} \bigg) \chi(z) =0.
\label{425}
\end{equation}

This is the parabolic cylinder equation \cite{abramowitz,zwilinger}
and the same equation appears in the work by D.T. Pegg, S.N. Barnett
{\it et al} on the minimum uncertainty states of the angular momentum and the
angular position (see \cite{pegg1} Eq. (37)). The general solution
of (\ref{425}) is well known \cite{abramowitz} and using formulas (\ref{422})
and (\ref{423}) one gets the general solution of (\ref{421}) with
$\Delta n \not=0$ as
$$
\psi(\phi) = e^{-i \langle n \rangle \phi} e^{-{1 \over 2}{\Delta n \over
\sqrt{\langle \phi^2 \rangle}} \phi^2} \bigg\{ a_{1} \cdot \ {_1}F_1
\bigg({1 \over 2} \cdot \bigg[{1\over 2} - \Delta n \cdot
\sqrt{\langle \phi^2 \rangle}\bigg], {1 \over 2}, {\Delta n \over
\sqrt{\langle \phi^2 \rangle}}\phi^2  \bigg)
$$
\begin{equation}
+ a_2 \sqrt{{2 \Delta n \over \sqrt{\langle \phi^2 \rangle}} } \phi
\cdot \ {_1}F_1 \bigg({1 \over 2} \cdot \bigg[{3\over 2} - \Delta n
\cdot \sqrt{\langle \phi^2 \rangle}\bigg], {3 \over 2}, {\Delta n
\over \sqrt{\langle \phi^2 \rangle}}\phi^2  \bigg) \bigg\},\; a_1, a_2 \in {\mathbb C},
\label{426}
\end{equation}
where $_1F_1$ stands for the confluent hypergeometric function.

In order to find a physically acceptable solution $\psi(\phi)$
we must extract from the general formula (\ref{426}) a function
which fulfills the periodicity conditions
\begin{equation}
\psi(\pi) = \psi(-\pi), \ \ \ \ \ \ \ {d \psi
(\pi) \over d \phi} = {d \psi (-\pi) \over d \phi}
\label{427}
\end{equation}
and has the Fourier expansion of the form (\ref{249}). Therefore
\begin{equation}
\int_{-\pi}^\pi \psi(\phi) e^{-i k \phi} d \phi =0, \ \ \ \ \ {\rm
for} \ \ \ \ k=1,2,\dots
\label{428}
\end{equation}
This last condition makes our problem drastically different from the
problem stated in \cite{pegg1}, where the minimum uncertainty product
states of the angular momentum and the angular position are studied.
Substituting  Eq. (\ref{426}) into (\ref{427}) and using the
abbreviations
\setcounter{orange}{1}
\renewcommand{\theequation} {\arabic{section}.\arabic{equation}\theorange}
\begin{equation}
\label{429a}
 y_1(\phi)=  e^{-{1 \over 2}{\Delta n \over \sqrt{\langle
\phi^2 \rangle}} \phi^2} \cdot  \ {_1}F_1 \bigg({1 \over 2} \cdot
\bigg[{1\over 2} - \Delta n \cdot \sqrt{\langle \phi^2
\rangle}\bigg], {1 \over 2}, {\Delta n \over \sqrt{\langle \phi^2
\rangle}}\phi^2 \bigg),
\end{equation}
\addtocounter{orange}{1}
\addtocounter{equation}{-1}
\begin{equation}
 y_2(\phi)=  e^{-{1 \over 2}{\Delta n \over \sqrt{\langle
\phi^2 \rangle}} \phi^2} \sqrt{{2 \Delta n \over \sqrt{\langle
\phi^2 \rangle}} } \phi \cdot  \ {_1}F_1 \bigg({1 \over 2} \cdot
\bigg[{3\over 2} - \Delta n \cdot \sqrt{\langle \phi^2
\rangle}\bigg], {3 \over 2}, {\Delta n \over \sqrt{\langle \phi^2
\rangle}}\phi^2 \bigg) \label{429b}
\end{equation}
\renewcommand{\theequation} {\arabic{section}.\arabic{equation}}
one gets the following system of equations (remember that
$y_1(-\phi)=y_1(\phi)$ and $y_2(-\phi) = -y_2(\phi)$)
\setcounter{orange}{1}
\renewcommand{\theequation} {\arabic{section}.\arabic{equation}\theorange}
\begin{equation}
\label{430a}
a_2 \cos \big(\langle n \rangle \pi\big) \cdot y_2(\pi) = ia_1 \sin
\big(\langle n \rangle \pi\big) \cdot y_1(\pi),
\end{equation}
\addtocounter{orange}{1}
\addtocounter{equation}{-1}
\begin{equation}
a_1 \cos \big(\langle n \rangle \pi\big) \cdot {dy_1(\pi) \over d
\phi} = ia_2 \sin \big(\langle n \rangle \pi\big) \cdot  {dy_2(\pi)
\over d \phi}.
\label{430b}
\end{equation}
\renewcommand{\theequation} {\arabic{section}.\arabic{equation}}
The Wro\'nski determinant $W(y_1(\phi),y_2(\phi))$ is given by
$$
W(y_1(\phi),y_2(\phi)) = y_1(\phi) {dy_2(\phi) \over d \phi} -
y_2(\phi) {dy_1(\phi) \over d \phi} = {\rm const.}
$$
\begin{equation}
= y_1(0) {dy_2(0) \over d \phi} = \bigg({2 \Delta n \over
\sqrt{\langle \phi^2 \rangle}} \bigg)^{1/2}.
\label{431}
\end{equation}
Without any loss of generality we can put  $a_1$ real
\begin{equation}
a_1^* = a_1.
\label{432}
\end{equation}

Assuming that the wave function $ \psi(\phi)$ given by
(\ref{426}) is normalized to $1$, and taking into account Eqs.
(\ref{429a}), (\ref{429b}) and (\ref{432}) one quickly finds
\begin{equation}
\langle n \rangle = \int_{-\pi}^\pi \psi^*(\phi) i{\partial \over
\partial \phi} \psi(\phi) d \phi
= \langle n \rangle + i \int_{-\pi}^\pi a_1 \cdot \bigg( a_2
y_1(\phi){dy_2(\phi) \over d \phi} + a_2^* y_2(\phi){dy_1(\phi)
\over d \phi} \bigg)   d\phi.
\end{equation}
Hence
\begin{equation}
a_1 \cdot \int_{-\pi}^\pi \bigg( a_2 y_1(\phi){dy_2(\phi) \over d
\phi} + a_2^* y_2(\phi){dy_1(\phi) \over d \phi} \bigg)   d\phi = 0.
\label{434}
\end{equation}
Multiplying the Wro\'nskian (\ref{431}) by $a_1 a_2$, 
integrating out over $d \phi$ and comparing with Eq.
(\ref{434}) we get
\begin{equation}
a_1 \cdot \int_{-\pi}^\pi ( a_2 + a_2^*) y_2(\phi){dy_1(\phi) \over
d \phi}  d\phi = - 2 \pi a_1 a_2   \bigg({2 \Delta n \over
\sqrt{\langle \phi^2 \rangle}}\bigg)^{1/2}.
\end{equation}

Hence, the product $a_1 a_2$ is a real number. Concluding, 
without any loss of generality one can put the coefficients $a_1$
and $a_2$ real i.e.
\begin{equation}\label{436}
\chi^*(\phi) = \chi(\phi)
\end{equation}
with $\chi(\phi)$ defined by (\ref{422}).

Then returning to the periodicity conditions (\ref{430a}), (\ref{430b}) we easily
realize that the analysis of these conditions splits into three
cases:
\begin{enumerate}
\item
\label{i1}
  $a_1 y_1(\pi) =0 = a_2y_2(\pi)$ and $a_1 {d
y_1(\pi) \over d \phi} = 0 = a_2 {d y_2(\pi) \over d \phi}$,
\item 
\label{i2} $\sin (\langle n \rangle \pi) =0,$ $a_2 y_2(\pi)
=0$ and  $a_1 {d y_1(\pi) \over d \phi} = 0$,
\item 
\label{i3}
  $\cos (\langle n \rangle \pi) =0,$ $a_1 y_1(\pi)
=0$ and  $a_2 {d y_2(\pi) \over d \phi} = 0$.
\end{enumerate}

Consider the case \ref{i1}. Here one immediately infers that from the
theorem on existence and uniqueness of a solution of the initial value
problem for Eq. (\ref{423}) it follows that the unique solution of
this equation fulfilling the conditions posed in \ref{i1}  is
$\chi(\phi)=0$. So $\psi(\phi)$ is also equal to zero and such a
wave function does not represent any quantum state.

In the case \ref{i2} we find that $\langle n \rangle=N$,
$N=0,1,2,\dots$. Hence $\psi(\phi)= e^{-iN \phi} \chi(\phi),$
$N=0,1,2,\dots$ Inserting this $\psi(\phi)$ into (\ref{428}),
taking then the complex conjugate of both sides and employing
(\ref{436}) one obtains
\begin{equation}
\int_{-\pi}^\pi \chi(\phi) e^{i(N+k)\phi} d \phi   = \int_{-\pi}^\pi
\psi(\phi) e^{i(2N+k)\phi} d \phi =0, \ \ \ k=1,2, \dots
\label{437}
\end{equation}

Finally, by (\ref{428}) and (\ref{437}) we conclude that in the case
\ref{i2} the respective wave function must be of the form
\begin{equation}
 \psi(\phi) = {1 \over \sqrt{2 \pi}} \sum_{n=0}^{2N} c_n
e^{-in \phi}.
\label{438}
\end{equation}

Substituting (\ref{438}) into Eq. (\ref{421}), taking the first and
then the second derivative of both sides and comparing the values of
left hand sides at $\phi=\pi$ and $\phi=-\pi$ one gets that if
$\Delta n \not=0$ then
\begin{equation}
\psi(\pi) =0 \ \ \ \ \ {\rm and} \ \ \ \ \ {d \psi (\pi) \over d
\phi} =0.
\end{equation}
Consequently, as in the preceding case \ref{i1}, by the uniqueness of
the solution of the initial value problem for Eq. (\ref{421}), the
only solution for the case \ref{i2} is $\psi(\phi) =0$. Finally, in
the case \ref{i3} one has $\langle n \rangle = N+{1 \over 2}$,
$N=0,1,2,\dots $ So $ \psi(\phi)= e^{-iN\phi} e^{-{i \over 2}
\phi} \chi(\phi)$. Inserting this wave function into (\ref{428}) and
taking the complex conjugate of the integral obtained, using also
(\ref{436}) we get
\begin{equation}
\int_{-\pi}^\pi \chi(\phi) e^{iN \phi} e^{{i \over 2}\phi}e^{ik\phi}
d \phi = \int_{-\pi}^\pi \psi(\phi) e^{i(2N+1+k)\phi} d \phi =0, \ \
\ k=1,2, \dots
\label{440}
\end{equation}
From (\ref{428}) and (\ref{440}) it follows that in the case \ref{i3}
the wave function is of the form
\begin{equation}
 \psi(\phi) = {1 \over \sqrt{2 \pi}} \sum_{n=0}^{2N+1} c_n
e^{-in \phi}.
\end{equation}
Consequently, the analogous arguments as in the case \ref{i2} lead to
the conclusion that the only solution of the case \ref{i3} is
$\psi(\phi)=0$.

Gathering our rather long discussion we find that: {\it there is no
minimum uncertainty product state for the quantum phase and the number
of photons if $\Delta n \not= 0$.} Thus one arrives at the following

\begin{tw}
\label{t4.2}
{\rm
The only minimum uncertainty product states for the  quantum phase and the
number of photons are the eigenstates $|n \rangle$, $n=0,1, \dots ,$
of the photon number operator $\widehat{n}$.

The eigenstates  $|n \rangle$ are the global minimum uncertainty product
states for the quantum phase and the number of photons. }\rule{2mm}{2mm}
\end{tw}

From this theorem we get an important 

\begin{co}
\label{co4.2}
{\rm
Let $|\psi \rangle$, $\langle \psi | \psi \rangle =1$, be any state
which is not an eigenstate of the photon number operator
$\widehat{n}$, i.e. $\Delta n \not =0$ and let the product of
uncertainties $\Delta n$ and $\widetilde{\Delta \phi}$ in  $|\psi
\rangle$ be $\Delta n \cdot \widetilde{\Delta \phi} = b$. Then in
any neighborhood of $|\psi \rangle$ there exists a state $|\psi '
\rangle$, $\langle \psi' | \psi' \rangle =1$, such that the product
of uncertainties  $(\Delta n)'$ and $(\widetilde{\Delta \phi})'$ in
$|\psi ' \rangle$ is less than $b$ i.e. $(\Delta n)' \cdot
(\widetilde{\Delta \phi})' < \Delta n \cdot \widetilde{\Delta \phi}
=b$.} \rule{2mm}{2mm}
\end{co}
\noindent
(For the definition of $\widetilde{\Delta \phi}$ see
Example \ref{ex2.1}).

\section{The Minimum Uncertainty Sum States}

\setcounter{ex}{0}
\setcounter{tw}{0}
\setcounter{co}{0}
\setcounter{equation}{0}

In this section we are going to find the states which {\it minimize
locally} the sum of uncertainties  $(\Delta f_1)^2 + (\Delta
f_2)^2$. This problem reduces to the isoperimetric variational
problem
$$
\delta {\cal G}[\psi(\phi),\psi^*(\phi)] =0, \ \ \ \ \ \ \ \ \ \ \int_{-\pi}^\pi
\psi^*(\phi) \psi(\phi)  d \phi =1,
$$
where
$$
{\cal G}[\psi(\phi),\psi^*(\phi)] = (\Delta f_1)^2 + (\Delta f_2)^2
$$
\begin{equation}
= \int_{-\pi}^\pi \bigg(\delta \widehat{F}_1 \psi(\phi)\bigg)^*
\delta \widehat{F}_1 \psi(\phi) d \phi  +  \int_{-\pi}^\pi
\bigg(\delta \widehat{F}_2 \psi(\phi)\bigg)^* \delta \widehat{F}_2.
\psi(\phi) d \phi, \label{51}
\end{equation}
It  leads to the following equation
\begin{equation}
\bigg[(\delta \widehat{F}_1)^\dag \delta \widehat{F}_1 + (\delta
\widehat{F}_2)^\dag \delta \widehat{F}_2  - \tau \bigg] \psi(\phi) =0,
\label{52}
\end{equation}
where $\tau \in \mathbb{R}$ is a Lagrange multiplier. Multiplying
(\ref{52}) by $\psi^*(\phi)$ and integrating over $d \phi$ we obtain that
$\tau = (\Delta f_1)^2 + (\Delta f_2)^2$. Finally, Eq. (\ref{52})
reads
\begin{equation}
\bigg\{(\delta \widehat{F}_1)^\dag \delta \widehat{F}_1 + (\delta
\widehat{F}_2)^\dag \delta \widehat{F}_2 - \bigg[(\Delta f_1)^2 +
(\Delta f_2)^2 \bigg] \bigg\} \psi(\phi) =0.
\label{53}
\end{equation}
Note that for $\Delta f_1 = \Delta f_2$ Eqs. (\ref{generaleqn45})
and (\ref{53}) are equivalent.

As the first example we consider

\begin{ex}
\label{ex5.1}
\end{ex}
Here we take $f_1=f_1(\phi)= e^{-i \phi}$ and $f_2 =f_2(n) = n.$

Then Eq. (\ref{53}) gives now
\begin{equation}
\bigg\{ \bigg( i {\partial \over \partial \phi} - \langle n \rangle
\bigg)^2 - \big\langle e^{i \phi} \big\rangle e^{-i \phi} -
\big\langle e^{- i \phi} \big\rangle e^{i \phi} + \bigg(2
\big\langle e^{i \phi} \big\rangle \big\langle e^{- i \phi}
\big\rangle - (\Delta n)^2  \bigg) \bigg\} \psi(\phi) =0.
\label{54}
\end{equation}
Assume that $\Delta n \not =0$. Then one quickly shows that the
results of our considerations done in \ref{i147} and \ref{i248} of Example
\ref{ex4.1} hold true if we change the `product of uncertainties'  to
the `sum of uncertainties'. Thus, for example, one should change
(\ref{producto416}) to
\begin{equation}
(\Delta n)'^2 + (\Delta e^{-i \phi})'^2 < (\Delta n)^2 + (\Delta
e^{-i \phi})^2
\end{equation}
and (\ref{desigprime418}) to
\begin{equation}
(\Delta n)'^2 + (\Delta e^{-i \phi})'^2 > (\Delta n)^2 + (\Delta
e^{-i \phi})^2.
\end{equation}
It remains only to study the case when $\Delta n =0$. Now we have of
course $\psi(\phi) ={1 \over \sqrt{2 \pi}} e^{-in\phi}, $ $n = 0,1,
\dots ,$ $\langle n \rangle = n$, $\langle e^{-i \phi} \rangle =0$
and $\Delta e^{-i \phi} =1.$ So
\begin{equation}
(\Delta n)^2 + (\Delta e^{-i \phi})^2 =1.
\label{57}
\end{equation}
The question is,  if the states $\psi(\phi) ={1 \over
\sqrt{2 \pi}} e^{-in\phi}$, $n = 0,1, \dots ,$  are really the
minimum uncertainty sum states for the phase function$e^{-i \phi}$
 and the number of photons.

To answer this question let us consider the following state
\begin{equation}
\psi''(\phi):={1 \over \sqrt{2 \pi}} \bigg(\sqrt{1 -
\varepsilon}e^{-in \phi} + \sqrt{\varepsilon} e^{-i(n + 1)
\phi}\bigg), \ \ \ \ \ \ \ 0< \varepsilon <1. \label{58}
\end{equation}
Simple calculations lead to the following results
\setcounter{orange}{1}
\renewcommand{\theequation} {\arabic{section}.\arabic{equation}\theorange}
\begin{equation}
(\Delta n)''^2 = \varepsilon \cdot (1 - \varepsilon),
\end{equation}
\addtocounter{orange}{1}
\addtocounter{equation}{-1}
\begin{equation}
\big\langle e^{-i \phi} \big\rangle''=  \sqrt{\varepsilon \cdot (1 -
\varepsilon)},
\end{equation}
\addtocounter{orange}{1}
\addtocounter{equation}{-1}
\begin{equation}
\big(\Delta e^{-i\phi} \big)''^2 = 1 - \big| \big\langle e^{-i \phi}
\big\rangle''  \big|^2 = 1 - \varepsilon \cdot (1 - \varepsilon).
\label{59}
\end{equation}
\renewcommand{\theequation} {\arabic{section}.\arabic{equation}}
Hence
\begin{equation}
(\Delta n)''^2 +  \big(\Delta e^{-i\phi} \big)''^2 =1,
\label{510}
\end{equation}
as in (\ref{57}). However, the function (\ref{58}) does not satisfy
Eq. (\ref{54}) for any $0 < \varepsilon < 1.$ Therefore the state
$\psi ''(\phi)$ given by (\ref{58}) is not a (local) minimum uncertainty
sum state. On the other hand $\lim_{\varepsilon \to 0} \psi''(\phi) = {1 \over
\sqrt{2 \pi}} e^{-in\phi}$ and the sum (\ref{510}) of uncertainties for
$\psi''(\phi)$  is equal to the sum  (\ref{57}) of the
uncertainties for $ \psi(\phi) = {1 \over \sqrt{2 \pi}}
e^{-in\phi}.$ Since the state $\psi''(\phi)$ is not a
minimum uncertainty sum state, for any $0 < \varepsilon < 1$ in an
arbitrary neighborhood  of $\psi''(\phi)$ (in the sense of the norm
$\|\cdot \|_0$) there exists a state such that the respective sum of
uncertainties is less than 1. Consequently, in any neighborhood of
the state   $\psi(\phi) = {1 \over \sqrt{2 \pi}} e^{-in\phi}$ lies a
state for which this sum is less than that given in (\ref{57}) i.e.
1. This means that the states  $ \psi(\phi) = {1 \over \sqrt{2 \pi}}
e^{-in\phi},$ $n=0,1,\dots  $ are not the minimum uncertainty sum
states.

Gathering the results obtained in the present example one
arrives at the conclusion.

\begin{tw} \label{t5.1}
{\rm
There are no minimum uncertainty sum states for the phase function
$e^{-i\phi}$ and the number of photons $n$. }\rule{2mm}{2mm}
\end{tw}

 It is an easy matter to show that the analogous theorems hold
true in the cases of phase functions $e^{i\phi}$, $\cos \phi$ or
$\sin \phi$ and the number of photons $n$.

Theorem \ref{t5.1} leads immediately to the following 

\begin{co}
 \label{co5.1}
{\rm
Let $|\psi \rangle$, $\langle \psi | \psi \rangle =1$, be any state,
and let $\Delta n$ and $\Delta e^{-i \phi}$ are the uncertainties of
the number of photons $n$ and the phase function $e^{-i \phi}$,
respectively in $|\psi \rangle$. Then in any neighborhood of $|\psi
\rangle$ (in the sense of the norm $\| \cdot \|_0$) there exists a
state $|\psi' \rangle$, $\langle \psi' | \psi' \rangle =1$, such
that $(\Delta n)'^{\, 2} +  (\Delta e^{-i \phi})'^{\, 2} < (\Delta n)^2 +
(\Delta e^{-i \phi})^2,$ where $(\Delta n)'$ and  $(\Delta e^{-i
\phi})'$ stand for the respective uncertainties in  $|\psi'
\rangle$.

The same holds true in the cases of phase functions $e^{i \phi}$,
$\cos \phi$ or $\sin \phi$ and the number of photons $n$.} \rule{2mm}{2mm}
\end{co}
We end the considerations of this section with an important 

\begin{ex} \label{ex5.2}
\end{ex}
We are looking now for the minimum uncertainty sum states for the
quantum phase and the number of photons.

Employing the results of Examples 2.1 and 4.2 one finds that in the
present case Eq. (\ref{53}) reads
\begin{equation}
\bigg\{\bigg(i {\partial \over \partial \phi} -\langle n \rangle
\bigg)^2 + \phi^2 - \bigg[\langle \phi^2 \rangle + (\Delta n)^2
\bigg] \bigg\} \psi(\phi) =0.
\label{511}
\end{equation}
We quickly recognize that Eq. (\ref{511}) can be obtained from
(\ref{421}) by substitutions ${(\Delta n)^2 \over \langle \phi^2
\rangle} \to 1$ and $2(\Delta n)^2 \to  \langle \phi^2 \rangle +
(\Delta n)^2.$ So we define $ \chi(\phi)$ as in (\ref{422})
and, according to (\ref{424}), the variable
\begin{equation}
x = \sqrt{2}\phi.
\end{equation}
Then Eq. (\ref{511}) reduces to the parabolic cylinder equation
analogous to (\ref{425})
\begin{equation}
{d^2 \chi(x) \over d x^2} - \bigg({x^2 \over 4} - {\langle \phi^2
\rangle +  (\Delta n)^2 \over 2} \bigg) \chi(x) =0.
\end{equation}
Finally, the general solution of Eq. (\ref{511}) reads (compare with
(\ref{426}))
$$
\psi(\phi) = e^{-i \langle n \rangle \phi} e^{-{1 \over 2}\phi^2} \bigg\{
a_{1} \cdot \  {_1}F_1 \bigg({1 \over 2} \cdot \bigg[{1\over 2} -
{\langle \phi^2 \rangle +  (\Delta n)^2 \over 2} \bigg], {1 \over
2}, \phi^2 \bigg)
$$
\begin{equation}
+ a_2 \sqrt{2} \phi  \cdot \  {_1}F_1 \bigg({1 \over 2} \cdot
\bigg[{3\over 2} - {\langle \phi^2 \rangle +  (\Delta n)^2 \over 2}
\bigg], {3 \over 2}, \phi^2 \bigg)\bigg\}, \ \ \ \ \ a_1,a_2 \in
\mathbb{C}.
\end{equation}
The further analysis is, {\it mutatis mutandi} the same as in
Example \ref{ex4.2}. Thus one arrives at the following 

\begin{tw}
\label{t5.2}
{\rm
There is no minimum uncertainty sum state for the quantum phase and the
number of photons.} \rule{2mm}{2mm}
\end{tw}
From this theorem we get also an important 
\begin{co}
\label{co5.2}
{\rm
For any state $|\psi \rangle$, $\langle \psi | \psi \rangle =1$, and
for any neighbourhood of $|\psi \rangle$ (in the sense of the norm
$\| \cdot \|_0$) there exists a state $|\psi' \rangle$, $\langle
\psi' | \psi' \rangle =1$, such that $(\Delta n)'^{\,2} +
(\widetilde{\Delta \phi})'^{\,2} < (\Delta n)^2 + (\widetilde{\Delta
\phi})^2$. }\rule{2mm}{2mm}
\end{co}

\section{Concluding Remarks}

Comparing our results with the ones concerning intelligent states
and the minimum uncertainty product or sum states for the angular momentum
and the angular position \cite{pegg1,gonzalez,kowalski} one quickly
notes that  most of the states, which have been found in
those papers and which  play an important role in quantum
mechanics on the circle, are not admitted in quantum optics. Of
course, the reason of this lies in the fact that, in contrast to
angular momentum $L_z$ of the particle on the circle, which can
assume values $L_z=0,\pm \hbar,\pm 2 \hbar, \dots,$ the number of
photons $n$ can be only a natural number $n=0,1,2,\dots$. So in
quantum optics the wave function $\psi(\phi)$ must satisfy the
condition (\ref{249}) which, as we have seen in the present paper, is
highly restrictive. In particular one can see this from Theorems \ref{t2.1}
and \ref{t4.2} which prove that the only number phase
Robertson -- Schr\"odinger and Heisenberg -- Robertson intelligent states
are the eigenstates $|n \rangle$, $n=0,1,2 \dots$ of the photon
number operator $\widehat{n}$ and the same states are also the only
minimum uncertainty product states for the quantum phase and the number of
photons. We can succinctly state that {\it  the photon is intelligent}.


\vskip 1.5truecm

\centerline{\bf Acknowledgments}

We are indebted to P. Brzykcy, M. Dobrski and T. Zawadzki for their
interest in this work. The work of H. G.-C. and F. J. T. was
partially supported by SNI-M\'exico and CONACyT research grants: 128761
and 103478. In addition F. J. T. was partially supported by COFAA-IPN
and by SIP-IPN grants 20141498 and 20150975.




\begin{thebibliography}{99}

\bibitem{heisenberg} W. Heisenberg,  Z. Phys.  {\bf 43} (1927) 172.

\bibitem{kennard} E.H. Kennard, Z. Phys.  {\bf 44} (1973) 326.

\bibitem{weyl} H. Weyl,  {\it Gruppentheorie und Quantenmechanik} (Hirzel Verlag,
Leipzig, 1928).

\bibitem{robertson1} H.P. Robertson, Phys. Rev.  {\bf 34} (1929) 163.

\bibitem{robertson2} H.P. Robertson, Phys. Rev.  {\bf 35} (1930) 667A.

\bibitem{schrodinger} E. Schr\"odinger, Sitz. Preus. Acad. Wiss.
(Phys.-Math. Klasse) {\bf 19} (1930) 296.

\bibitem{robertson3} H.P. Robertson, Phys. Rev.  {\bf 46} (1934) 794.

\bibitem{trifonov1} D.A. Trifonov, J. Math. Phys. {\bf 34} (1993) 100.

\bibitem{trifonov2} D.A. Trifonov, J. Math. Phys. {\bf 35} (1994) 2297.

\bibitem{trifonov3} D.A. Trifonov, J. Phys. A: Math. Gen. {\bf 30} (1997) 5941.

\bibitem{trifonov4} D.A. Trifonov, Phys. Scripta {\bf 58} (1998) 246.

\bibitem{trifonov5} D.A. Trifonov, S.G. Donev, J. Phys. A: Math. Gen. {\bf 31} (1998) 8041.

\bibitem{trifonov6} D.A. Trifonov, J. Phys. A: Math. Gen. {\bf 33} (2000)
L299; J. Phys. A: Math. Gen. {\bf 34} (2001) L75.

\bibitem{trifonov7} D.A. Trifonov, in: {\it Geometry, Integrability and Quantization,
Varna 1999} (Coral Press Sci. Publ., Sofia 2000) pp. 257-282.

\bibitem{trifonov8} D.A. Trifonov, Eur. Phys. J. B {\bf 29} (2002) 349.

\bibitem{dodonov} V.V. Dodonov, E.V. Kurmyshev, V.I. Man'ko, Phys. Lett. A  {\bf 79} (1980) 150.

\bibitem{klauder} J.R. Klauder, B.S. Skagerstam, {\it Coherent States}, (World Scientific, Singapore
1985).

\bibitem{zang} W.-M. Zang, D.H. Feng, R. Gilmore,  Rev. Mod. Phys.  {\bf 62} (1990) 867.

\bibitem{tanas} R. Tana\'s,  {\it Lectures on Quantum Optics} (in polish),
\url{ http://zon8.physd.amu.edu.pl/~tanas}, 2007.

\bibitem{kryszewski} S. Kryszewski, Quantum Optics, Lecture notes for students,
\url{iftia9.uviv.gda.pl/~sjk/QO-SK.pdf}, Gda\'nsk 2009-2010.

\bibitem{IP1} L. Infeld, J. Pleba\'nski, Acta Phys. Polon. {\bf XIV} (1955) 41.

\bibitem{pleban} J. Pleba\'nski, Bull. Acad. Polon. Sci. Cl. {\bf III}, {\bf II} (1954) 213;
J. Pleba\'nski, Acta Phys. Polon. {\bf XIV} (1955) 275; J. Pleba\'nski, Phys. Rev. {\bf 101} (1956) 1825.

\bibitem{yuen}  H.P. Yuen, Phys. Rev. A {\bf 13} (1976) 2226.

\bibitem{loudon} R. Loudon, P. Knight, J. Mod. Opt. {\bf 34} (1987)
 709.

\bibitem{aragone} C. Aragone, E. Chalbaud, S. Salam, J. Math. Phys.
{\bf 17} (1976) 1963.

\bibitem{vaccaro} J.A. Vaccaro, D.T. Pegg, J. Mod. Opt. {\bf 37} (1990) 17.

\bibitem{PT} M Przanowski, F.J. Turrubiates, J. Phys. A: Math. Gen. {\bf 35} (2002) 10643.

\bibitem{smithhey} D.T. Smithey, M. Beck, J. Cooper, M.G. Raymer, Phys. Rev A {\bf 48} (1993) 3159.

\bibitem{brif1} C. Brif, Y. Ben --  Aryeh, Phys. Rev. A {\bf 50} (1994)
3505.

\bibitem{pegg1} D.T. Pegg, S.N. Barnett, R. Zambrini, S. Franke-Arnold, M. Padgett, New J. Phys.
{\bf 7} (2005) 62.

\bibitem{curthright} T. Curtright, C. Zachos, Mod. Phys. Lett. A {\bf 16}
(2001) 2381.

\bibitem{carruthers1} P. Carruthers, M.M. Nieto, Phys. Rev. Lett. {\bf 14} (1965)
387.

\bibitem{carruthers2} P. Carruthers, M.M. Nieto, Rev. Mod. Phys.  {\bf 40} (1968)
411.

\bibitem{jackiw} R. Jackiw, J. Math. Phys. {\bf 9} (1968) 339.

\bibitem{garrison} J.C. Garrison, J. Wong, J. Math. Phys. {\bf 11}
(1970) 2242.

\bibitem{levy} J.M. Levy -- Leblond, Ann. Phys. (N.Y.) {\bf 101} (1976)
319.

\bibitem{yamamoto} Y. Yamamoto, S. Machida, N. Imoto, M. Kitagawa,
G. Biork, J. Opt. Soc. Am. B {\bf 4} (1987) 1645.

\bibitem{luks1} A. Luk\v s, V. Pe\v rinov\'a, J. Kr\v epelka, Czech. J. Phys.
{\bf 42} (1992) 59.

\bibitem{luks2} A. Luk\v s, V. Pe\v rinov\'a, Phys. Rev. A {\bf 45} (1992)
6710.

\bibitem{brif2} C. Brif, Class. Quantum Grav. {\bf 12} (1995) 803.

\bibitem{mendas} I. Menda\v s, D.B. Popovi\'c, Phys. Rev. A {\bf 52} (1995) 4356.

\bibitem{luo} S. Luo, Phys. Lett. A {\bf 275} (2000) 165.

\bibitem{skagerstam} B.-S. K. Skagerstam, B.\AA. Bergsjordet, Phys.
Scripta {\bf 70} (2004) 26.

\bibitem{scharat} H.S. Sharatchandra, arXiv:quant-ph/9710020v1
(1997).

\bibitem{shapiro} J.H. Shapiro, S.R. Shepard, Phys. Rev. A {\bf 43}
(1991) 3795.

\bibitem{pereira} T. Pereira,  D.H.U. Marchetti, Prog. Theor. Phys.
{\bf 122} (2009) 1137.

\bibitem{susskind} L. Susskind, J. Glogower, Physics {\bf 1} (1964)
49.

\bibitem{pegg2} D.T. Pegg, S.M. Barnett, Europhys. Lett. {\bf 6}
(1988) 483.

\bibitem{pegg3} D.T. Pegg, S.M. Barnett, Phys. Rev. A {\bf 39} (1989)
1665.

\bibitem{barnett} S.M. Barnett, D.T. Pegg, J. Modern Opt. {\bf 36}
(1989) 7.

\bibitem{busch} P. Busch, M. Grabowski, P. Lahti, Ann. Phys. {\bf
237} (1995) 1.

\bibitem{przan1} M. Przanowski, P. Brzykcy, Ann. Phys.
{\bf 337} (2013) 34.

\bibitem{przan2} M. Przanowski, P. Brzykcy, J. Tosiek, Ann. Phys.
{\bf 351} (2014) 919.

\bibitem{perinova} V. Pe\v rinov\'a, A. Luk\v s, J. Pe\v rina, {\it Phase in
Optics}, (World Scientific, Singapore, 1998).

\bibitem{miranowicz}
A. Miranowicz, {\it Lectures on Quantum Optics} (in polish),
\url{ http://zon8.physd.amu.edu.pl/~miran}, 2008.

\bibitem{octavio}
O. Casta\~{n}os, R. L\'{o}pez -- Pe\~{n}a, M. A. Man'ko and V. Man'ko, {\it Squeezing Operator and Squeeze Tomography}, in {\it Topics in Mathematical Physics, General Relativity and Cosmology in Honor of Jerzy Pleba\'nski}, . H.  Garc\'{\i}a -- Compe\'an, B. Mielnik, M. Montesinos, M. Przanowski (Eds.), (World Scientific, London, 2006) 109.


\bibitem{london} F. London, Z. Phys. {\bf 37} (1926) 915; Z. Phys.
{\bf 40} (1927) 193.

\bibitem{luis} A. Luis, L.L. S\'anchez -- Soto, Phys. Rev. A {\bf 48}
(1993) 752.

\bibitem{biatynicka} Z. Bia{\l}ynicka -- Birula, I. Bia{\l}ynicki -- Birula, J.
Appl. Phys. B {\bf 60} (1995) 275.

\bibitem{gerhardt} H. Gerhardt, H. Welling, D. Fr\"olich, Appl.
Phys. {\bf 2} (1973) 91.

\bibitem{nieto} M.M. Nieto, Phys. Lett. A {\bf 60} (1977) 401.

\bibitem{lyrich} R. Lynch, Phys. Rev. A {\bf 41} (1990) 2841.

\bibitem{gerry} C.C. Gerry, K.E. Urba\'nski, Phys. Rev. A {\bf 42}
(1990) 662.

\bibitem{gantsog} T.S. Gantsog, A. Miranowicz, R. Tana\'s, Phys.
Rev. A {\bf 46} (1992) 2870.

\bibitem{judge} D. Judge, Il Nuovo Cim. {\bf XXXI} (1964) 332.

\bibitem{luks3} A. Luk\v s, V. Pe\v rinov\'a, Czech. J. Phys. {\bf 41}
(1991) 1205.

\bibitem{luks4} A. Luk\v s, V. Pe\v rinov\'a, Phys. Scripta T {\bf 48} (1993)
94.

\bibitem{abramowitz} A. Abramowitz, I.A. Stegun, (Eds.) {\it Handbook
of Mathematical Functions with Formulas, Graphs, and Mathematical
Tables}, (Dover, New York, 1972).

\bibitem{zwilinger} D. Zwillinger, {\it Handbook of Differential
Equations,} (Academic Press, Boston, 1997).

\bibitem{gonzalez} J.A. Gonz\'alez, M.A. del Olmo, J. Phys. A: Math.
Gen. {\bf 31} (1998) 8841.

\bibitem{kowalski} K. Kowalski, J. Rembieli\'nski, J. Phys. A: Math.
Gen. {\bf 35} (2002) 1405.


\end{thebibliography}
\end{document}